\documentclass[aps,amsmath,twocolumn,floatfix,superscriptaddress,prb,footinbib,10pt]{revtex4-2}

\usepackage{amsmath,amsfonts,amssymb,amscd,bm,mathtools,mathrsfs,comment}
\usepackage{setspace}[10]
\usepackage[italicdiff]{physics}
\usepackage{siunitx}
\usepackage{graphicx}
\usepackage[usenames, dvipsnames]{color}
\usepackage{xcolor}
\usepackage{tabularray}
\PassOptionsToPackage{hyphens}{url}%url wrapping
\usepackage[hyperfootnotes=false]{hyperref}
\hypersetup{% hyperref
  setpagesize=false,
  bookmarksnumbered=true,%
  bookmarksopen=true,%
  colorlinks=true,%
  linkcolor=blue,
  citecolor=blue,
  urlcolor=blue,
}

\usepackage{stmaryrd}
\usepackage{mathrsfs}
\usepackage{braket}
\usepackage{xspace}
\usepackage{comment}
\usepackage{bm}
\usepackage{bbm}
\definecolor{lightblue}{rgb}{0.13, 0.26, 0.99}

\usepackage{xurl}

\usepackage{mathtools}

\usepackage[normalem]{ulem}

\usepackage[version=3]{mhchem}

\usepackage{empheq}

\newcommand{\Unit}[2]{#1\,\unit{#2}}

\allowdisplaybreaks

\begin{document}

\title{Floquet Theory for Light-Driven Rotation of Dipolar and Multipolar Particles}

\author{Amane Takano}
\email{a.takano@chiba-u.jp}
\affiliation{Department of Physics, Chiba University, Chiba 263-8522, Japan}

\author{Minoru Kanega}
\email{m.kanega.phys@chiba-u.jp}
\affiliation{Department of Physics, Chiba University, Chiba 263-8522, Japan}

\author{Masahiro Sato}
\email{sato.phys@chiba-u.jp}
\affiliation{Department of Physics, Chiba University, Chiba 263-8522, Japan}

\date{\today}
%##########################################################

\begin{abstract} % Up to 600 characters.
  Nano- or micro-particle rotation driven by light has been well known in the fields of optical manipulation and optical physics since the end of the last century. It is viewed as a sort of angular-momentum transfer from light to material, but its microscopic analysis based on the Hamiltonian or the equation of motion has been less developed.
  We model this rotation with a simple setup of an electrically dipolar or multipolar particle irradiated by circularly polarized laser and comprehensively analyze the Langevin-type equation of motion by using the Floquet theory for dissipative classical systems and the mode separation method. Furthermore, we numerically compute the time evolution of the particle. As a result, we accurately estimate the dependence of the laser-frequency, laser-intensity, particle mass, temperature, and friction (dissipation) on the laser-driven rotation. We determine the ``nonequilibrium phase diagram'' of the laser-driven rotation in a broad parameter regime, which consists of three regimes: the rotation frequency $\Omega\propto\omega^{-1}$, $\Omega\propto\omega^{-3}$, or $\Omega=\omega$ ($\omega$ is the laser frequency). Comparing our theoretical result with some experiments, we show that the result of the overdamped Langevin equation is qualitatively consistent with the experiments.
\end{abstract}
\maketitle

\begin{figure}[tb]
  \centering
  \includegraphics[width=\linewidth]{"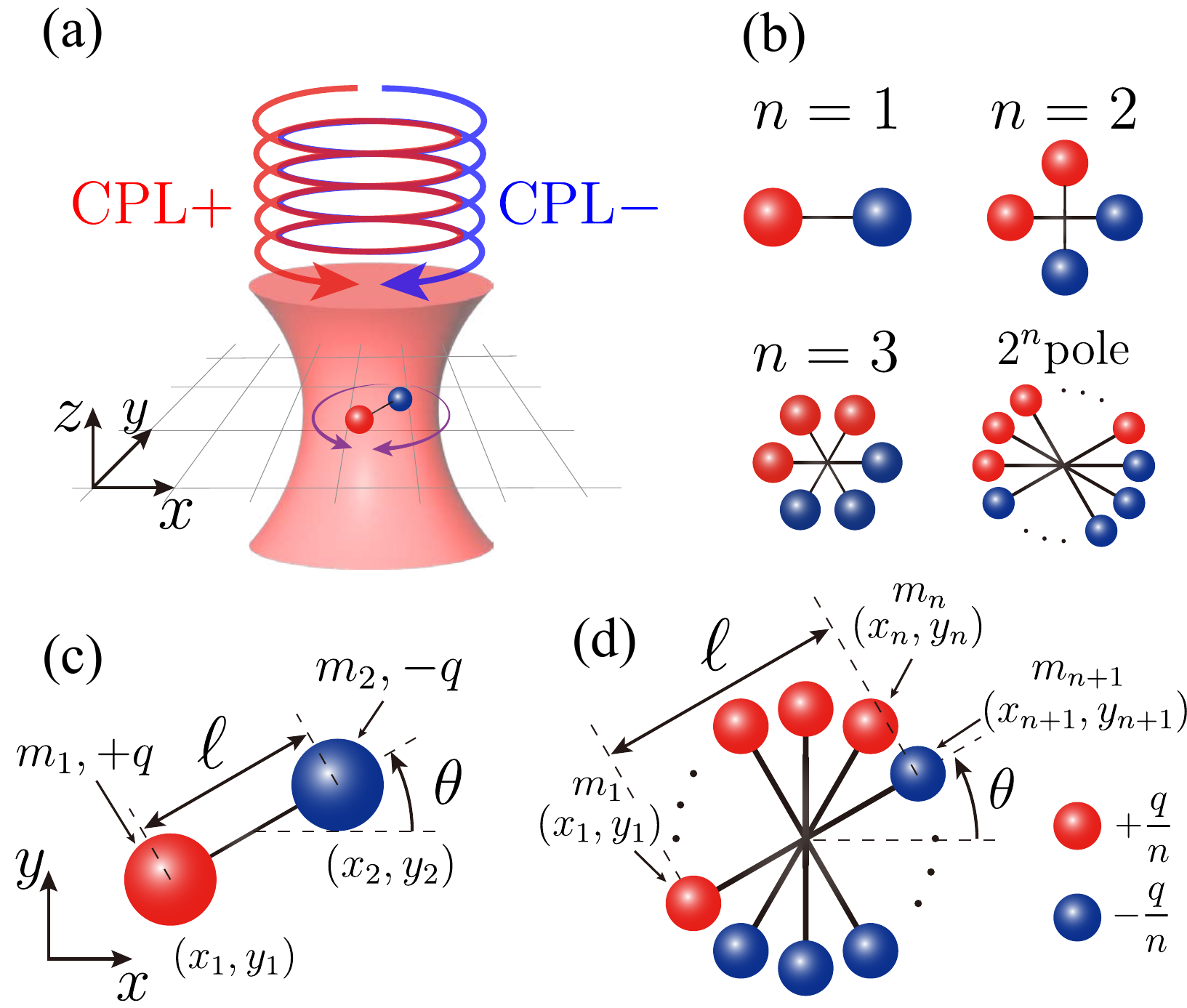"}
  \caption{
    (a) Schematic of the setup: radiation-pressure trapping and a multipolar model confined to two dimensions, driven by right-/left-handed CPL (magnetic field neglected).
    (b) Electric multipolar ($2^{n}$-polar) model.
    (c) Schematic of the electric dipole: masses $m_{1},m_{2}$, charges $+q,-q$, and rotation angle $\theta$.
    (d) Schematic of the electric multipole.
  }
  \label{fig:model}
\end{figure}

%%%%%%%%%%%%%%%%%%%%%%%%%%%%%%%%%%%%%%%%%%%%%%%
%%%%%%%%%%%%%%%%%%%%%%%%%%%%%%%%%%%%%%%%%%%%%%%
%%%%%%%%%%%%%%%%%%%%%%%%%%%%%%%%%%%%%%%%%%%%%%%
\textit{Introduction}---%
Optical manipulation techniques, especially optical tweezers [Fig.~\ref{fig:model}(a)], enable contactless trapping and control of submicro- to micro-sized particles and provide a versatile means of probing their microscopic properties~\cite{ashkin1970acceleration,ashkin1971optical,ashkin1987optical-science,ashkin1987optical-nature,ishijima2001single,bustamante2003ten,kobayashi2025formation}.
Experiments using circularly polarized light (CPL)~\cite{friese1998optical,tong2010alignment,lehmuskero2013ultrafast,reimann2018ghz} and optical vortices carrying orbital angular momentum (OAM)~\cite{tao2005fractional,dienerowitz2008optical,shen2016trapping,chen2019optical} have shown that micro particles can rotate at frequencies far below the optical drive, commonly interpreted as angular-momentum transfer from the beam~\cite{beth1936mechanical,allen1992orbital,franke2008advances}.
The driving frequencies are typically in the visible-to-near-infrared range, $f=\omega/(2\pi)\sim10^{14}\text{--}10^{15}\,\unit{Hz}$, whereas the rotational response is usually $f_{\mathrm{rot}}=\Omega/(2\pi)\sim\unit{Hz}\text{--}\unit{kHz}$~\cite{friese1998optical,tong2010alignment,lehmuskero2013ultrafast}; even in vacuum, where GHz rotation has been demonstrated~\cite{reimann2018ghz}, the pronounced scale separation $f_{\mathrm{rot}}\ll f$ still holds.
Theoretical research for this rotating phenomenon has been actively performed and often tailored the analysis of optical torque, a specific coupling mechanism, or material settings~\cite{tong2010alignment,lehmuskero2013ultrafast,tamura2019interparticle,hoshina2020nanoscale}.
Nevertheless, there has been no microscopic theory for solving the equation of motion (EOM) or predicting the rotation speed so far.

%%%%%%%%%%%%%%%%%%%%%%%%%%%%%%%%%%%%%%%%%%%%%%%
%%%%%%%%%%%%%%%%%%%%%%%%%%%%%%%%%%%%%%%%%%%%%%%

The fact of $f_{\mathrm{rot}}\ll f$ implies that the light driven rotation may be interpreted as a Floquet engineering phenomenon.
Floquet engineering on time-periodic drives has become a well-established route to controlling nonequilibrium responses and emergent properties on time scales distinct from the drive period~\cite{oka2009photovoltaic,kitagawa2011transport,goldman2014periodically,bukov2015universal,eckardt2015high,sato2016laser,mori2016rigorous,kuwahara2016floquet,mikami2016brillouin,hubener2017creating,eckardt2017colloquium,oka2019floquet,sato2021floquet,ivanov2021floquet,tanaka2024theory}.
Based on Floquet theory~\cite{shirley1965solution,sambe1973steady}, many advances have been made in quantum systems, including driven electrons in solids~\cite{oka2009photovoltaic,kitagawa2011transport,mikami2016brillouin,hubener2017creating,tanaka2024theory} and spin systems~\cite{sato2016laser,sato2021floquet}.

While the motion of micro-particles is expected to be adequately described by classical physics, Floquet theory has been developed to focus on quantum systems over the last decade.
In addition, Floquet theory has typically been applied to periodic drive systems where the time scale difference between the periodic external field and the motion of interest is at most about $10^2-10^3$.
However, as mentioned above, $f$ is extremely larger than $f_{\rm rot}$ ($f\sim 10^6-10^{14}f_{\rm rot}$) in the case of laser-driven particle rotation. It is generally challenging to derive a relation between such two different time-scale motions even if one uses the Floquet theory. These facts would be the main reason why the quantitative analysis of light-driven rotation has not been performed so far.

In this paper, we analyze the CPL-driven rotation based on the Langevin EOMs for a fundamental model of laser-irradiated micro-particles.
To this end, we use the Floquet theory for classical stochastic systems, which has recently been developed by our group~\cite{higashikawa2018floquet,sato2025floquet}.
Through the high-frequency (HF) expansion for the models, we derive effective EOMs describing the slow rotation (revolution) in both the underdamped and overdamped regimes.
This yields the parameter (laser-frequency $\omega$, laser-intensity $E_0$, particle mass, friction, etc.) dependence of the CPL-driven rotation frequency $\Omega$ in analytic fashion: In the high-frequency regime $\Omega\ll \omega$, we obtain $\Omega\propto E_0^2\omega^{-3}$ for the underdamped system and $\Omega\propto E_0^2\omega^{-1}$ for the overdamped system.
We refer to the resulting slow rotation as ``Floquet rotation''.
In low-frequency regime, on the other hand, we obtain an ac-field-following rotation with $\Omega=\omega$ via the numerical calculation of the EOMs.
Combining the analytical result and comprehensive numerical calculation of Langevin EOMs, we complete the nonequilibrium steady-state (NESS) ``phase diagram'' for CPL-driven rotation in a broad parameter range, which possesses three regions: $\Omega\propto E_0^2\omega^{-3}$, $\Omega\propto E_0^2\omega^{-1}$, and $\Omega=\omega$.
Our result can be compared with future experiments and provides a benchmark for quantitative analyses of optical manipulation.

%%%%%%%%%%%%%%%%%%%%%%%%%%%%%%%%%%%%%%%%%%%%%%%%%%%%%%%%%%%%%
%%%%%%%%%%%%%%%%%%%%%%%%%%%%%%%%%%%%%%%%%%%%%%%%%%%%%%%%%%%%%
%%%%%%%%%%%%%%%%%%%%%%%%%%%%%%%%%%%%%%%%%%%%%%%%%%%%%%%%%%%%%
\textit{Model}---%
To capture the essence of the CPL-driven rotation, we consider the setup that a rigid assembly of point charges trapped in two-dimensional (2D) space is irradiated by CPL as shown in Fig.~\ref{fig:model}(a).
Hereafter, we use ``multipolar model'' or ``multipole'' to represent the rigid charged particles shown in Fig.~\ref{fig:model}(b).
As shown in Fig.~\ref{fig:model}(c) and (d), the electric $2^{n}$-polar assembly is composed of $n$ electric dipoles ($2n$ point particles), and its diameter is $\ell$. The $i$-th particle at $\bm{r}_{i}=(x_{i},y_{i})$ has a mass $m_{i}$ and a charge $q_{i}$ for $i\in\{1,2,\ldots,2n\}$ [see Fig.~\ref{fig:model}(c) for the dipole case]. The charge distribution is defined as $q_i = q/n$ and $q_{n+i} = -q/n$ for $i\in\{1,2,\ldots,n\}$, and the rigid body is totally charge neutral. Under the assumption that the rigid multipole irradiated by a CPL is in a certain medium (liquid water, air, etc.), the classical EOM for the $i$-th particle is given by
\begin{align}
  m_{i}\ddot{\bm{r}}_{i} &= -\gamma_n\dot{\bm{r}}_{i}
  + q_{i}\bm{E}
  -\kappa_n\bm{r}_{i}
  + \bm{F}^{(i)} + \bm{h}^{(i)},
  \label{eq:ULE-multipole}
\end{align}
where $\gamma_n = \gamma/n$ is the viscous drag coefficient of the medium, and $\bm{h}^{(i)}=(h^{(i)}_{x}(t),h^{(i)}_{y}(t))$ represents the thermal fluctuation force, modeled as Gaussian white noise satisfying the fluctuation-dissipation relation.
The electric field $\bm{E}(t)$ of the applied CPL is defined by
\begin{align}
  \bm{E}(t)=E_{0}(\cos\omega t,\,\sin\omega t)
  \label{eq:Efield}
\end{align}
with $\omega$ being the angular frequency. We have assumed that the CPL also induces the trap potential $U(\bm{r})=\kappa_n \bm{r}^{2}/2$ with $\kappa_n=\kappa/n$ and the force $-\kappa_n\bm{r}_{i}$~\cite{iida2008theory}.
In the real experiments, two fields $\bm E(t)$ and $U(\bm{r})$ are correlated with each other as they stem from the applied CPL. However, for simplicity, we consider them as two independent parameters and will concentrate on the role of the electric field $\bm E(t)$. The force $\bm{F}^{(i)}=(F_{x}^{(i)},F_{y}^{(i)})$ denotes the internal force enforcing the rigid-body constraints: $\sum_{i}F_{x}^{(i)} = \sum_{i}F_{y}^{(i)} = 0$. Following the terminology of nonequilibrium physics, we should call Eq.~\eqref{eq:ULE-multipole} an underdamped Langevin equation (ULE).
See the Supplemental Material (SM)~\cite{SuppMaterial} for the normalization of $\gamma_n,\kappa_n$ and the noise correlations.
Since we consider the particle rotation in the trapped 2D space, we introduce the polar coordinate and define the relative coordinate angle $\theta$ as shown in Fig.~\ref{fig:model} (d).

When the rigid multipole follows a rotating motion in the application of CPL, the magnitudes of velocity $|\dot{\bm{r}}_{i}|$ and acceleration $|\ddot{\bm{r}}_{i}|$ are roughly estimated as $|\dot{\bm{r}}_{i}|\sim \ell \Omega$ and $|\ddot{\bm{r}}_{i}|\sim \ell\Omega^2$.
If the condition $m_i \ell\Omega^2\ll \gamma_n \ell \Omega$ ($m_i|\ddot{\bm{r}}_{i}|\ll \gamma_n |\dot{\bm{r}}_{i}|$) holds in an experiment, the inertial term to the slow motion with the frequency $\Omega$ is negligible
\footnote{Here, we assume that when we judge whether a CPL-driven multipole belongs to the ULE or OLE, the CPL-driven fast micromotion is irrelevant and the time-averaged rotation $\Omega$ is relevant.}.
In this case, the EOM may be reduced to the following overdamped Langevin equation (OLE)
\begin{align}
  \gamma_n\dot{\bm{r}}_{i}  &= q_{i}\bm{E} -\kappa_n\bm{r}_{i} + \bm{F}^{(i)} + \bm{h}^{(i)}. \label{eq:OLE-multipole}
\end{align}
In the rest of the main text, we will focus mainly on the zero-temperature $T=0$ cases of both the ULE and OLE because we can show that the temperature effect is negligible when $T$ is at most room temperature. 
Details are provided in SM~\cite{SuppMaterial}.

\begin{figure}[tb]
  \centering
  \includegraphics[width=\linewidth]{"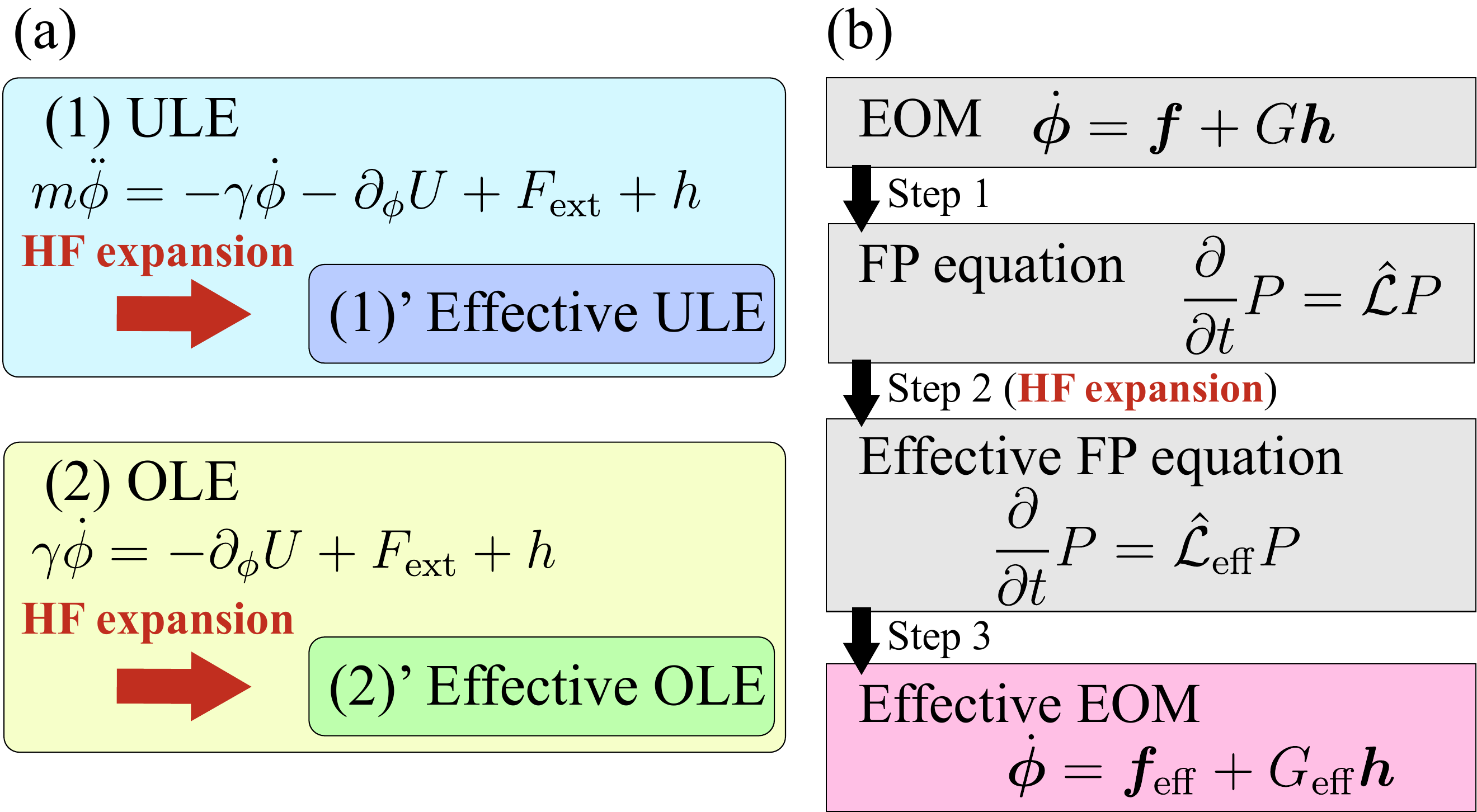"}
  \caption{
    (a) Underdamped Langevin equation (ULE) (1), overdamped Langevin equation (OLE) (2), and the corresponding effective EOMs (1)' and (2)' obtained via HF expansion based on Floquet theory~\cite{higashikawa2018floquet,sato2025floquet}.
    Symbols $\phi$, $m\ddot{\phi}$, $-\gamma\dot{\phi}$, $U(\phi)$, $F_{\mathrm{ext}}(\phi,t)$, and $h(t)$ are respectively the classical variable, the inertial term, the friction, the potential, a periodic driving term, and the thermal-noise term.
    (b) Logical flow of Floquet HF expansion approach in Refs.~\cite{higashikawa2018floquet,sato2025floquet}.
    Since the Langevin EOM is generally nonlinear in terms of $\phi$ and contains stochastic noise, the Floquet theorem cannot be directly applied.
    Instead, we consider the Fokker--Planck (FP) equation for the probability distribution function $P(\phi)$.
    The first line $\dot{\bm\phi}=\bm{f}+G\bm h$ denotes the matrix form of the EOM such as Eqs.~(1) and (2) in panel (a). Here, $\bm{f}(\bm{\phi},t)$ and $G(\bm{\phi},t)$ are the time-periodic deterministic force and noise-amplitude matrix, respectively, and $\hat{\mathcal{L}}(\bm{\phi},t)$ is the corresponding time-periodic FP operator.
    The HF expansion gives the time-independent effective quantities $\hat{\mathcal{L}}_{\mathrm{eff}}(\bm{\phi})$, $\bm{f}_{
    \mathrm{eff}}(\bm{\phi})$, and $G_{\mathrm{eff}}(\bm{\phi})$.
  }
  \label{fig:equations}
\end{figure}

%%%%%%%%%%%%%%%%%%%%%%%%%%%%%%%%%%%%%%%%%%%%%%%%%%%%%%%%%%%%%
%%%%%%%%%%%%%%%%%%%%%%%%%%%%%%%%%%%%%%%%%%%%%%%%%%%%%%%%%%%%%
%%%%%%%%%%%%%%%%%%%%%%%%%%%%%%%%%%%%%%%%%%%%%%%%%%%%%%%%%%%%%
\textit{EOM in HF Regimes}---%
In the following, we consider the high-frequency (HF) regimes of the ULE [Eq.~\eqref{eq:ULE-multipole}] and the OLE [Eq.~\eqref{eq:OLE-multipole}].
One can apply the standard Floquet expansion to periodically driven quantum systems in the HF regime, whereas the Floquet theorem cannot be applied directly to classical systems because the EOM is generally nonlinear in terms of mechanical variables $\phi$. However, we have recently developed the Floquet theory for classical systems [see Fig.~\ref{fig:equations} (b)].
Instead of solving the EOM of the mechanical variables, if we consider the EOM of distribution function $P(\phi)$, that is usually referred to as the Fokker-Planck equation (FPE), the Floquet theorem is applicable to the FPE because of its linearity in terms of $P(\phi)$. Using the Floquet expansion to the FPE, one obtains the effective FPE describing the slow dynamics.
From the effective FPE, we can return to the effective EOM for $\phi$. Following this formalism of Fig.~\ref{fig:equations} (b)~\cite{higashikawa2018floquet,sato2025floquet}, we obtain the effective EOMs for the OLE and ULE. 
Further details are provided in SM~\cite{SuppMaterial}.

%##########################################################
%%%%%%%%%%%%%%%%%%%%%%%%%%%%%%%%%%%%%%%%%%%%%%%%%%%%%%%%%%%%%%%%

For the OLE of the dipolar ($n=1$) model, we find that at $T=0$, where the thermal-noise contribution is negligible, the steady state rotates uniformly as $\dot{\theta}\simeq\Omega_{1}^{\mathrm{O}} + \mathcal{O}(\omega^{-3})$, where
\begin{align}
  \Omega_{1}^{\mathrm{O}}\coloneq\frac{2}{\omega}\qty(\frac{qE_{0}}{\gamma \ell})^2,
  \label{eq:eff-OLE-dipole-3variables-Omega}
\end{align}
denotes the time-averaged angular velocity of the angle $\theta$ [see Fig.~\ref{fig:model}(c)].
This indicates that the optical electric field drives a slow rotation with angular velocity $\Omega_{1}^{\mathrm{O}}\propto\omega^{-1}\ll\omega$ in the HF regime. In the regime, the fast oscillation of $\bm E(t)$ averages out, leaving a net second-order ($\propto E_0^2$) effective torque and thereby producing the Floquet rotation (see Supplemental Movies~1 and~2~\cite{SuppMaterial}).
For the multipolar models, we obtain an analogous Floquet rotation with the same parameter dependence, $\Omega_{n}^{\mathrm{O}}\propto(qE_{0}/(\gamma \ell))^{2}\omega^{-1}$, as shown in Table~\ref{tab:frequency}.

The HF expansion can also be carried out for the ULE, leading to an effective ULE.
For the dipolar ($n=1$) model with $m_1=m_2$, we obtain the solution
$\dot{\theta}\simeq\Omega_{1}^{\mathrm{U}}+\mathcal{O}(\omega^{-5})$, in which the leading term is given by
\begin{align}
  \Omega_{1}^{\mathrm{U}}\coloneq\frac{1}{2\omega^{3}}\qty(\frac{qE_{0}}{\mu_{1}\ell})^{2},
  \label{eq:omega_ULE}
\end{align}
where $\mu_{1}=m_{1}m_{2}/(m_{1}+m_{2})$ is the reduced mass.

In the mass-imbalance case of $m_1\neq m_2$, an orbital rotation can occur in addition to the autorotation (spin) that we have considered so far.
The former stands for rotation of the center-of-mass coordinate.
In the HF regime, we find a steady-state solution in which the spin and orbital motions co-rotate with a common time-averaged angular velocity $\Omega_{\mathrm{rot}}$.
For $\abs{\delta} \ll 1$, it is given by
\begin{align}
\Omega_{\mathrm{rot}}=\Omega_{1}^{\mathrm{U}}\left(1+\delta+\mathcal{O}(\delta^{2})\right)
  \label{eq:omega_rot}
\end{align}
where $\delta\propto (m_1-m_2)^2$ is the dimensionless parameter defined in SM~\cite{SuppMaterial}. For the ULE of multipolar models with $m_1=m_2=\cdots=m_{2n}$, one can easily derive $\Omega_{n}^{\mathrm{U}}\propto(qE_{0}/(\mu_{n}\ell))^{2}\omega^{-3}$, where $\mu_n$ is an effective mass parameter whose definition is given in SM~\cite{SuppMaterial}: For $m_1=m_2=\cdots=m_{2n}$, $\mu_n$ reduces to $M/4$ with $M=\sum_{j=1}^{2n} m_j$ being the total mass.

%%%%%%%%%%%%%%%%%%%%%%%%%%%%%%%%%%%%%
\renewcommand{\arraystretch}{1.15}
\begin{table}[tb]
  \caption{
    Prefactors $A_{n}^{\mathrm{O}}$ and $A_{n}^{\mathrm{U}}$ of CPL-driven rotation predicted by HF expansion.
    The angular velocities of the rotation are given by $\Omega_{n}^{\mathrm{O}}=A_{n}^{\mathrm{O}}(qE_{0}/(\gamma \ell))^{2}\omega^{-1}$ for the OLE and $\Omega_{n}^{\mathrm{U}}=A_{n}^{\mathrm{U}}(qE_{0}/(\mu_{n}\ell))^{2}\omega^{-3}$ for the ULE.
    Here $a_{n} \coloneq \csc^{2}\qty(\frac{\pi}{2n})$, and $a_{n}/n^{2}\to 4/\pi^{2}$ as $n\to\infty$.
    The explicit form of $\mu_{n}$ is given in SM~\cite{SuppMaterial}. In the case of ULE, we set $m_1=m_2=\cdots=m_{2n}$.
  }
  \label{tab:frequency}
  \begin{ruledtabular}
    \begin{tabular}{cccccc}
      & $n=1$ & $n=2$ & $n=3$ & $n$ & $n\to\infty$ \\
      \hline
      $A_{n}^{\mathrm{O}}$ & $2$ & $1$ & $8/9$ & $2a_{n}/n^{2}$ & $8/\pi^{2}$ \\
      $A_{n}^{\mathrm{U}}$ & $1/2$ & $1/4$ & $2/9$ & $a_{n}/(2n^{2})$ & $2/\pi^{2}$ \\
    \end{tabular}
  \end{ruledtabular}
\end{table}
%%%%%%%%%%%%%%%%%%%%%%%%%%%%%%%%%%%%%

Here, we briefly comment on the mode separation method that is another analytical method for periodically driven classical systems. Its essence is as follows: We first divide the classical variable of interest into the fast and slow components like $\theta=\theta_{\mathrm{fast}}+\theta_{\mathrm{slow}}$, secondly take the time average for the fast component, and finally obtain the EOM for the slow component. This is a heuristic method, but it is often powerful. This mode separation method derives the same results as Eqs.~\eqref{eq:eff-OLE-dipole-3variables-Omega} and \eqref{eq:omega_ULE} when the point-particle masses are all equal, $m_1=m_2=\cdots=m_{2n}$. However, it is difficult to extract the EOM for $\theta_{\rm fast}$ when there is a mass difference such as $m_1\neq m_2$. This indicates that the HF expansion in Fig.~\ref{fig:equations} (b) is applicable to a broad class of classical systems, including complicated ones.

%%%%%%%%%%%%%%%%%%%%%%%%%%%%%%%%%%%%%%%%%%%
\textit{Low-frequency Regimes}
Next, we consider the low-frequency (LF) regimes. Using the numerical calculation of both the OLE and ULE at $T=0$, we find $\dot{\theta}=\Omega=\omega$ in the LF regime.
Namely, the multipole completely follows the direction of the ac electric field $\bm E(t)$ at each moment in time. This is naturally understood because the multipole tends to decrease the light-matter coupling energy.

%%%%%%%%%%%%%%%%%%%%%%%%%%%%%%%%%%%%%%%%%%%%%%%%%%%%%%%
\begin{figure}[tb]
  \centering
  \includegraphics[width=\linewidth]{"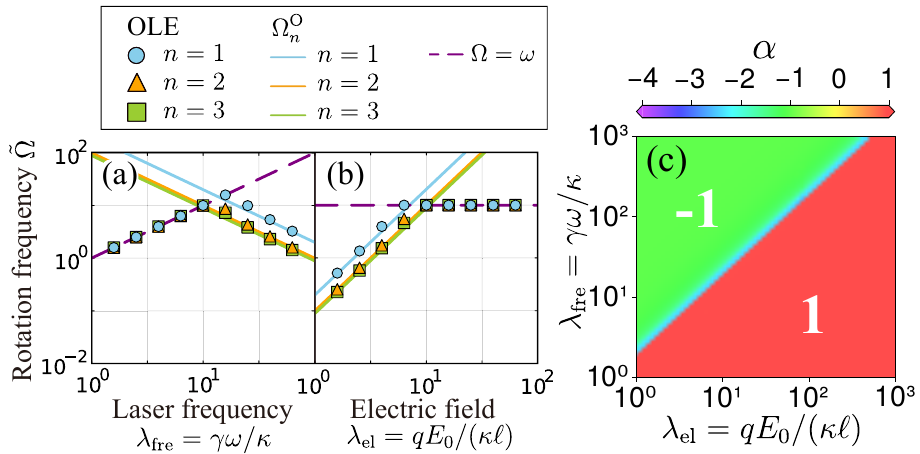"}
  \caption{
    Panels (a) and (b) show the CPL-driven rotation frequency in the NESS of the OLE at $T=0$, $\tilde{\Omega}=\gamma\Omega/\kappa$, as functions of $\lambda_{\mathrm{fre}}$ at $\lambda_{\mathrm{el}}=10$ and of $\lambda_{\mathrm{el}}$ at $\lambda_{\mathrm{fre}}=10$, respectively.
    Points (circles, triangles, and squares) denote the numerical OLE result for the models with $n=1,2,3$, while solid lines show the analytical HF prediction $\Omega_{n}^{\mathrm{O}}$ and dashed lines indicate the field-following response $\Omega=\omega$.
    Panel (c) shows the ``phase'' diagram of NESS at $T=0$ in the $(\lambda_{\mathrm{el}},\lambda_{\mathrm{fre}})$ plane, that shows the exponent $\alpha$ of $\Omega\propto\omega^{\alpha}$; the green and red regions correspond to the $\alpha=-1$ and $\alpha=1$ regimes, respectively.
  }
  \label{fig:phase_diagram_ole}
\end{figure}
%%%%%%%%%%%%%%%%%%%%%%%%%%%%%%%%%%%%%%%%%%%%%%%%%

%%%%%%%%%%%%%%%%%%%%%%%%%%%%%%%%%%%%%%%%%%
\textit{Phase Diagram of NESS}---%
Based on these analytical and numerical results, we draw the phase diagram of the CPL-driven rotation in the NESS.
First, we consider the multipolar model following the OLE.
We numerically estimate the time-averaged angular velocity $\Omega$ at T=0.
The details of the numerical procedure are explained in SM~\cite{SuppMaterial}.
We introduce the dimensionless parameters $\lambda_{\mathrm{fre}}=\gamma\omega/\kappa$ and $\lambda_{\mathrm{el}}=qE_{0}/(\kappa \ell)$.
Panels (a) and (b) in Fig.~\ref{fig:phase_diagram_ole} show the parameter dependence of $\Omega$ at $T=0$. They indicate that the analytical prediction from the HF expansion is in good agreement with the numerically exact result. Panel (c) is the phase diagram of the NESS of the dipolar model ($n=1$) in the space $(\lambda_{\mathrm{el}},\lambda_{\mathrm{fre}})$ at $T=0$, which shows the exponent $\alpha$ of $\Omega\propto\omega^{\alpha}$.
The green and red regimes correspond to the Floquet rotation with $\alpha=-1$ and the field-following rotation of $\alpha=1$ ($\Omega=\omega$), respectively. As expected, the $\alpha=-1$ phase is expanded in the high-frequency (large $\lambda_{\mathrm{fre}}$) regime, while $\Omega=\omega$ holds in the low-frequency regime.

\begin{figure}[tb]
  \centering
  \includegraphics[width=\linewidth]{"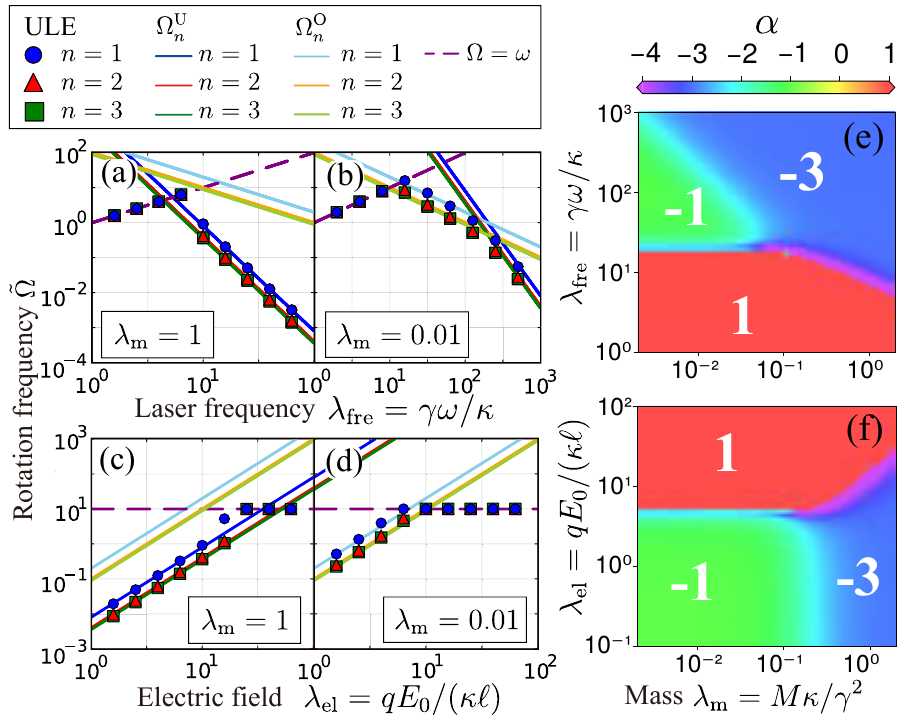"}
  \caption{
    Panels (a) and (b) [(c) and (d)] show the CPL-driven rotation frequencies as functions of $\lambda_{\mathrm{fre}}$ [$\lambda_{\mathrm{el}}$] for $\lambda_{\mathrm{m}}=1$ and $0.01$, respectively, at $\lambda_{\mathrm{el}}=10$ [$\lambda_{\mathrm{fre}}=10$] in the ULE of the multipolar models.
    Points (circles, triangles, and squares) show the numerical ULE results for multipoles with $n=1,2,3$; the colored solid lines denote $\Omega_{n}^{\mathrm{U}}$ and $\Omega_{n}^{\mathrm{O}}$, while the purple dashed line denotes $\Omega=\omega$. 
    The phase boundaries between the neighboring regimes with different $\alpha$ gradually shift with increasing $n$ due to the prefactors in Table~\ref{tab:frequency}.
    Panels (e) and (f) respectively show the phase diagrams of the NESS of the dipolar models at $T=0$ in the $(\lambda_{\mathrm{m}},\lambda_{\mathrm{fre}})$ space with $\lambda_{\mathrm{el}}=10$ and the $(\lambda_{\mathrm{m}},\lambda_{\mathrm{el}})$ space with $\lambda_{\mathrm{fre}}=10$. Blue, green, and red phases correspond to the $\alpha=-3$, $\alpha=-1$, and $\alpha=1$ regimes, respectively.
  }
  \label{fig:phase_diagram_ule}
\end{figure}

Next, we move to the CPL-driven rotation in the NESS of the ULE. We first focus on the equal-mass case of $m_1=m_2=\cdots=m_{2n}$.  Since the model has an inertial term, we introduce an additional dimensionless parameter $\lambda_{\mathrm{m}}=M\kappa/\gamma^{2}$.
In the high-frequency regime, Eq.~\eqref{eq:omega_ULE} would hold if the inertial term is relevant, whereas
Eq.~\eqref{eq:eff-OLE-dipole-3variables-Omega} is expected to appear when the mass term is irrelevant. In fact, the mode separation method leads to both results of Eqs.~(\ref{eq:eff-OLE-dipole-3variables-Omega}) and (\ref{eq:omega_ULE}) in the ULE with the equal mass condition. Therefore, three ``phases'' are expected to appear in the ULE;
$\Omega\propto\omega^{-3}$, $\Omega\propto\omega^{-1}$, and $\Omega=\omega$.
Panels (a), (b), (c) and (d) of Fig.~\ref{fig:phase_diagram_ule} are the numerically computed CPL-driven rotation frequencies in the ULE for the multipolar models with $n=1,2,3$. They indicate that the numerical exact results agree well with the analytical predictions of Eqs.~(\ref{eq:eff-OLE-dipole-3variables-Omega}) and (\ref{eq:omega_ULE}). Panels (e) and (f) respectively show the phase diagrams of the NESS of the dipolar model ($n=1$) at $T=0$ in spaces $(\lambda_{\mathrm{m}},\lambda_{\mathrm{fre}})$ and $(\lambda_{\mathrm{m}},\lambda_{\mathrm{el}})$. As expected, blue ($\Omega\propto\omega^{-3}$), green ($\Omega\propto\omega^{-1}$) and red ($\Omega=\omega$) regimes appear.
In the higher-frequency (larger $\lambda_{\mathrm{fre}}$) or lower-field (smaller $\lambda_{\mathrm{el}}$) regimes, the Floquet rotation is realized.

%%%%%%%%%%%%%%%%%%%%%%%%%%%%%%%%%%%%%%%%%%%%%%
%##########################################################

\begin{figure}[tb]
  \centering
  \includegraphics[width=\linewidth]{"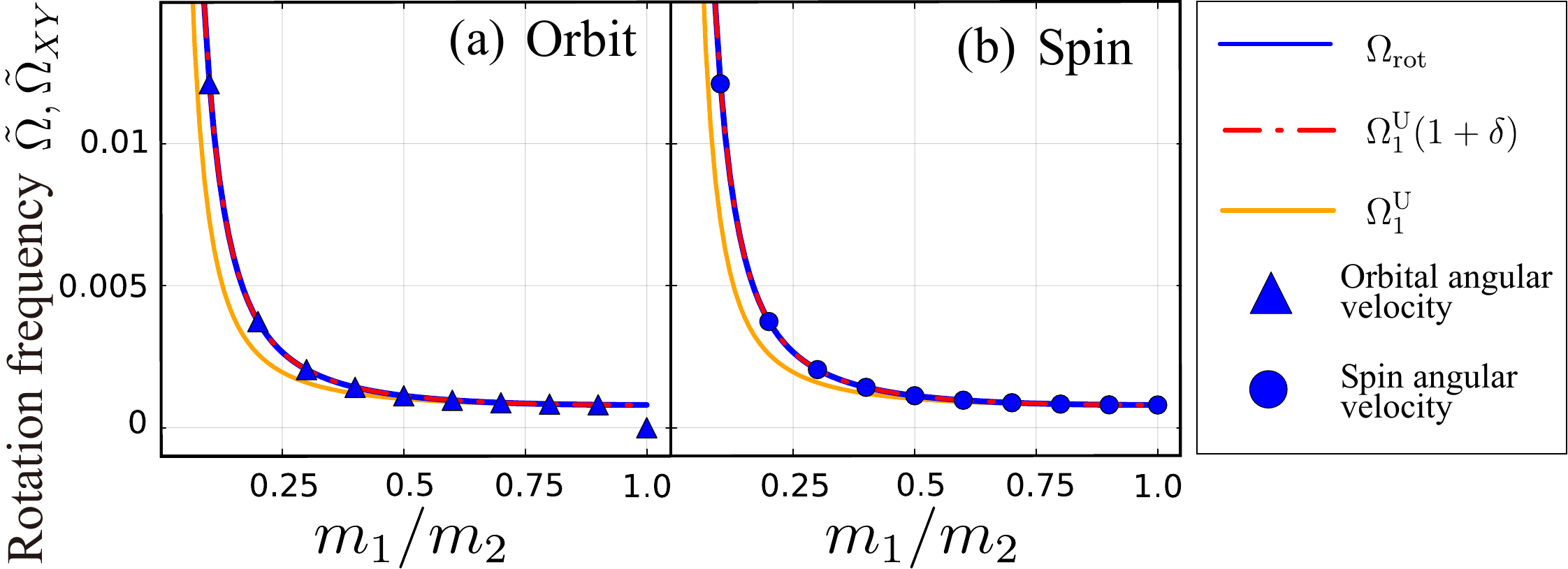"}
  \caption{
    Mass ratio $m_{1}/m_{2}$ dependence of the (a) orbital and (b) spin rotation frequencies, $\tilde{\Omega}_{XY}=\gamma\Omega_{XY}/\kappa$ and $\tilde{\Omega}=\gamma\Omega/\kappa$ (vertical axis) in the dipolar model at $\lambda_{\mathrm{m}}=1$, $\lambda_{\mathrm{fre}}=10^{2}$, $\lambda_{\mathrm{el}}=10$, and $T=0$.
    Points (triangles and circles) are the numerical results of ULE.
    The blue solid, red dash-dotted, and orange solid lines represent $\Omega_{\mathrm{rot}}$ given in Eq.~\eqref{eq:omega_rot}, its approximation $\Omega_{1}^{\mathrm{U}}(1+\delta)$, and $\Omega_{1}^{\mathrm{U}}$, respectively.
  }
  \label{fig:mass-imbalance}
\end{figure}

Let us turn to the case with different masses $m_{1}\neq m_{2}$ in the dipolar model.
In this case, as we mentioned, we have spin and orbital rotations.
Figure~\ref{fig:mass-imbalance} shows the dependence on the mass ratio $m_{1}/m_{2}$ of (a) the orbital angular velocity $\Omega_{XY}$ and (b) the spin angular velocity $\Omega$ in the high-frequency regime.
Panels (a) and (b) show that
the analytical HF expansion of Eq.~\eqref{eq:omega_rot} agrees with the numerically exact solutions with high precision. Moreover, the panels indicate that
mass difference is useful to enhance both the spin and orbital angular frequencies.

%%%%%%%%%%%%%%%%%%%%%%%%%%%%%%%%%%%%%%%
\textit{Finite temperature}---%
So far, we have mainly discussed the zero-temperature limit. Here we briefly comment on the temperature effect on the CPL-driven rotation.
In typical temperatures (such as room temperature) of the experimental studies of the CPL-driven rotation, we can numerically show that the rotational motion in the NESS is almost insensitive to temperature. 
Namely, the results of Figs.~\ref{fig:phase_diagram_ole} and \ref{fig:phase_diagram_ule} remain essentially unchanged at experimentally relevant temperatures within the present model.
For more detail, we discuss the temperature effects in SM~\cite{SuppMaterial}.

%%%%%%%%%%%%%%%%%%%%%%%%%%%%%%%%%%%%%%%%%%%%%%%%%%
%%%%%%%%%%%%%%%%%%%%%%%%%%%%%%%%%%%%%%%%%%%%%%%%%%

%%%%%%%%%%%%%%%%%%%%%%%%%%%%%%%%%%%%%%%%
%%%%%%%%%%%%%%%%%%%%%%%%%%%%%%%%%%%%%%%%
\textit{Comparison with experiments}---%
Finally, we compare our predictions with experiments.
Let us focus on the experiment of Ref.~\cite{tong2010alignment}, in which a cylinder-shaped Ag wire is irradiated by CPL.
Within our model, the wire is expected to be described by a dipolar model ($n=1$).
The wire's mass, length, and diameter are respectively estimated as $M\sim\Unit{400}{fg}$, $\ell\sim\Unit{5}{\mu m}$, and $\ell_{\rm d}\sim\Unit{100}{nm}$. The wavelength of CPL is $\Unit{830}{nm}$ ($\omega\simeq2\times\Unit{10^{15}}{rad/s}$) and the ac field at the focal point is $E_{0}\sim\Unit{10^{7}}{V/m}=\Unit{0.1}{MV/cm}$.
From Ref.~\cite{bonin2002light}, the friction factor $\gamma$ is estimated as $\gamma = \pi\eta \ell/(3\qty[\ln\qty(\ell/\ell_{\rm d})-0.66])$.
For water at 300--$\Unit{400}{K}$, the viscosity coefficient $\eta\sim0.22$--$\Unit{0.85}{mPa\cdot s}$, and thus $\gamma\sim(0.4$--1.4)$\Unit{\times10^{-9}}{kg/s}$.

The modeling of charge $q$ in the Ag wire is not easy.
For example, ferroelectrics such as \ce{BaTiO_3}~\cite{Uesu2016Ferroelectrics} can possess a polarization $\sim \Unit{10}{\mu C/cm^{2}}$. In this case, the edge charge is estimated as $q=PS\sim2\times\Unit{10^{-16}}{C}\sim1000e$, where $S=\pi\qty(\ell_{\rm d}/2)^{2}\sim8\times\Unit{10^{-15}}{m^{2}}$ is the cross section and $e$ is the elementary charge. Because Ag is metallic, a larger edge charge is expected to occur.
From this argument, if we set $q\sim 10^{4}$--$10^{6}e$, we obtain $\Omega_{1}^{\mathrm{U}}\sim10^{-19}$--$\Unit{10^{-16}}{rad/s}$ and $\Omega_{1}^{\mathrm{O}}\sim10^{-2}$--$\Unit{10^{3}}{rad/s}$.
As we mentioned in the Introduction, the observed rotation frequency is $10^{0}$--$10^3$ Hz, which is qualitatively compatible with the OLE-type estimate. In fact, for the Ag-wire parameters in the experiment of Ref.~\cite{tong2010alignment}, $M\Omega/\gamma\ll1$ holds~\footnotemark[1].
This supports the use of the OLE model as a phenomenological description of the observed time-averaged rotation.

%%%%%%%%%%%%%%%%%%%%%%%%%%%%%%%%%%%%
%%%%%%%%%%%%%%%%%%%%%%%%%%%%%%%%%%%%
%%%%%%%%%%%%%%%%%%%%%%%%%%%%%%%%%%%%
\textit{Conclusions}---%
We have developed a Floquet theory of the slow CPL-driven rotation in a minimal Langevin model of multipoles, yielding $\Omega\propto\omega^{-3}$ in the underdamped regime and $\Omega\propto\omega^{-1}$ in the overdamped regime.
Visible-light measurements should be compared at the parameter-exponent level, and a laser-frequency sweep provides the direct test of the exponents and crossover.
The present theory offers a benchmark against other optical torque contributions and identifies GHz/sub-GHz drives as a realistic window.
More broadly, it provides a concrete classical benchmark for testing Floquet-engineering ideas in periodically driven stochastic systems.

In the real experimental setup, the charge distribution of rotating micro particles would temporally change depending on the applied ac field. Therefore, developing a theory for such time dependent charge distributions is an intriguing future direction.

%%%%%%%%%%%%%%%%%%%%%%%%%%%%%%%%%%%%
%%%%%%%%%%%%%%%%%%%%%%%%%%%%%%%%%%%%
\textit{Acknowledgments}---%
M.S. is supported by JSPS KAKENHI (Grants No.~JP25K07198, No.~JP25H02112, No.~JP22H05131, No.~JP25H01609 and No.~JP25H01251) and JST, CREST Grant No.~JPMJCR24R5, Japan.

%%%%%%%%%%%%%%%%%%%%%%%%%%%%%%%%%%%%
%%%%%%%%%%%%%%%%%%%%%%%%%%%%%%%%%%%%

% \bibliography{main}
%apsrev4-2.bst 2019-01-14 (MD) hand-edited version of apsrev4-1.bst
%Control: key (0)
%Control: author (8) initials jnrlst
%Control: editor formatted (1) identically to author
%Control: production of article title (0) allowed
%Control: page (0) single
%Control: year (1) truncated
%Control: production of eprint (0) enabled
%

% \newpage

%######################################
%##### Supplementary Materials ########
%######################################

\setcounter{figure}{0}
\setcounter{equation}{0}
\setcounter{section}{0}

\onecolumngrid

\begin{center}

    \vspace{1.5cm}

    {\large \textbf{Supplemental Material: Floquet Theory for Light-Driven Rotation of Dipolar and Multipolar Particles}}

    \vspace{0.3cm}

    {\large Amane Takano${}^1$, Minoru~Kanega${}^1$, and~Masahiro~Sato$^1$} \\[2mm]
  \textit{${}^1$Department of Physics, Chiba University, Chiba 263-8522, Japan}\\

\end{center}

\vspace{0.6cm}

\renewcommand{\thesection}{}
\renewcommand{\thesubsection}{\Alph{subsection}}
\renewcommand{\theequation}{S\arabic{equation}}
\renewcommand{\thefigure}{S\arabic{figure}}
\renewcommand{\thetable}{S\arabic{table}}

%%%%%%%%%%%%%%%%%%%%%%%%%%%%%%%%%%%%%%%%%%
%%%%%%%%%%%%%%%%%%%%%%%%%%%%%%%%%%%%%%%%%%
%%%%%%%%%%%%%%%%%%%%%%%%%%%%%%%%%%%%%%%%%%

In this Supplementary Material, we discuss the details and techniques of our analysis used in the present study that are not explained carefully in the main text.

%%%%%%%%%%%%%%%%%%%%%%%%%%%%%%%%%%
\subsection{Langevin Equation, Thermal Noise, and Normalization}
Both the underdamped Langevin equation (ULE) and the overdamped Langevin equation (OLE) in the main text can be written in the form of a stochastic differential equation (SDE),
\begin{align}
  \dot{\bm{\phi}}(t)=\bm{f}(\bm{\phi},t)+G(\bm{\phi})\bm{h}(t).
  \label{eq:general-langevin}
\end{align}
Here, the vector $\bm{\phi}=(\phi_1,\phi_2,\cdots)^{\top}$ consists of the set of the independent coordinates ($4n$ components) for OLE, while it is the set of the coordinates and the corresponding velocities ($4n\times 2$ components) for the ULE. Since we consider the $2^n$-polar model in two-dimensional space, we have $2n\times 2$ coordinates.
The vector $\bm{f}$ describes the deterministic force, and the matrix $G$ connects the random-force vector $\bm{h}(t)=(h_1,h_2,\cdots)^{\top}=[h_x^{(1)},h_y^{(1)},\ldots,h_x^{(2n)},h_y^{(2n)}]^{\top}$ and the vector $\dot{\bm{\phi}}=(\dot{\phi}_1,\dot{\phi}_2,\cdots)^{\top}$.
Note that we assume that the drag (friction) is proportional to the velocities $\dot{\phi}_j$, while the accelerations $\ddot{\phi}_j$ are subject to the random force $h_j(t)$. The matrix $G$ is a square matrix for the OLE and is a rectangular type for the ULE.
For both models [see Eq.~(1) and (3) in the main text], we set the drag constants $\gamma_n=\gamma/n$ and the trap potential constant $\kappa_n=\kappa/n$ so that the total drag and restoring-force scales of the multipole remain independent of $n$.
Consistent with this normalization and the fluctuation--dissipation relation, the thermal random-force vector $\bm{h}^{(i)}=(h^{(i)}_{x}(t),h^{(i)}_{y}(t))$ is taken to be Gaussian white noise satisfying
\begin{align}
  \label{eq_Gauss_1}
  \langle h^{(i)}_{\alpha}(t)\rangle & = 0,\\
  \langle h^{(i)}_{\alpha}(t)h^{(j)}_{\beta}(t')\rangle
  & = (2\gamma k_{B}T/n)\delta_{ij}\delta_{\alpha\beta}\delta(t-t').
  \label{eq_Gauss_2}
\end{align}
Here, $i,j\in\{1,\ldots,2n\}$, $\alpha,\beta\in\{x,y\}$, $\langle\cdots\rangle$ denotes the ensemble average, $T$ is the environmental temperature, $k_{B}$ is the Boltzmann constant, and the corresponding diffusion constant is $D=\gamma k_{B}T/n$. We will use the dimensionless temperature
\begin{align}
  \lambda_{\mathrm{th}}& = k_{B}T/(\kappa\ell^{2}) \label{eq:lam_th}
\end{align}
in Sections~\ref{sec:NumericalMethod} and \ref{sec:thermal}.
In the zero-temperature limit considered in the analytical parts of the main text, $D=0$, so the stochastic term vanishes, and Eq.~\eqref{eq:general-langevin} reduces to the deterministic EOM $\dot{\bm{\phi}}=\bm{f}(\bm{\phi},t)$.

%%%%%%%%%%%%%%%%%%%%%%%%%%%%%%%%%%
\subsection{High-Frequency Expansion for Classical Equations of Motion}\label{sec:HF-expansion-general}
Equation~\eqref{eq:general-langevin} is generally nonlinear in $\bm{\phi}$ and stochastic, so the Floquet theorem cannot be applied to it directly. Following Refs.~\cite{higashikawa2018floquet,sato2025floquet}, we instead introduce the probability density $P(\bm{\phi},t)$, whose time evolution is governed by the time-periodic Fokker--Planck (FP) equation
\begin{align}
  \begin{aligned}
    \pdv{P}{t}
    &=\hat{\mathcal{L}}(\bm{\phi},t)P\\
    &=\sum_i\pdv{}{\phi_i}\qty[\mathcal{F}_i(\bm{\phi},t)P]
    +\sum_{i,j}\pdv[2]{}{\phi_i}{\phi_j}\qty[\mathcal{D}_{ij}(\bm{\phi})P].
  \end{aligned}
  \label{eq:general-fp}
\end{align}
For the SDE of Eq.~\eqref{eq:general-langevin}, $\mathcal{F}_i(\bm{\phi},t)$ and $\mathcal{D}_{ij}(\bm{\phi})$ are given by
\begin{align}
  \begin{aligned}
    \mathcal{F}_i(\bm{\phi},t)
    &=-f_i(\bm{\phi},t)-D\sum_{k,a}G_{ka}(\bm{\phi})
    \pdv{G_{ia}(\bm{\phi})}{\phi_k},\\
    \mathcal{D}_{ij}(\bm{\phi})
    &=D\sum_aG_{ia}(\bm{\phi})G_{ja}(\bm{\phi}).
  \end{aligned}
  \label{eq:general-fp-coefficients}
\end{align}
Equation~\eqref{eq:general-fp} is linear in $P$, and the periodic drive implies $\hat{\mathcal{L}}(\bm{\phi},t+\mathcal{T})=\hat{\mathcal{L}}(\bm{\phi},t)$ with $\mathcal{T}=2\pi/|\omega|$.
The HF expansion can therefore be applied to Eq.~\eqref{eq:general-fp}.
Using the Fourier decomposition
\begin{align}
  \hat{\mathcal{L}}(\bm{\phi},t)
  &=\sum_{m\in\mathbb{Z}}\hat{\mathcal{L}}_m(\bm{\phi})e^{-im\omega t},
\end{align}
the HF expansion yields the corresponding time-independent effective FP equation:
\begin{align}
  \pdv{P}{t}
  &\simeq\hat{\mathcal{L}}_{\mathrm{eff}}(\bm{\phi})P.
  \label{eq:effective-fp}
\end{align}
The periodic kick operator omitted in Eq.~\eqref{eq:effective-fp} describes the fast micromotion, whereas $\hat{\mathcal{L}}_{\mathrm{eff}}$ determines the slow motion and the time-averaged rotation frequency discussed in the present study.
When the truncated effective FP operator remains a second-order differential operator, it can be recast as an SDE with the time-independent quantities $\bm{f}_{\mathrm{eff}}$ and $G_{\mathrm{eff}}$ shown in Fig.~2(b) of the main text. It should be referred to as the Floquet effective equation of motion (EOM).
At finite temperature, the effective EOM can be represented in the form of a Markovian SDE up to the order of $\omega^{-2}$ when we start from a time dependent EOM with a Gaussian noise of Eq.~\eqref{eq_Gauss_1} and \eqref{eq_Gauss_2}.
However, if the HF expansion is on the order $\omega^{-3}$ or beyond, a non-Gaussian or non-Markovian noise may appear~\cite{sato2025floquet}.

The analytical expressions for the rotation frequencies in the main text are derived from the HF expansion at $T=0$.
In this limit, $D=0$, and the FP-operator expansion reduces to an expansion of the deterministic force.
If we use the Fourier representation $\bm{f}(\bm{\phi},t)=\sum_{m\in\mathbb{Z}}\bm{f}_m(\bm{\phi})e^{-im\omega t}$, where $\bm{f}_{-m}=\bm{f}_m^{*}$, the operator commutators used in the HF expansion are then expressed in terms of the Lie bracket
\begin{align}
  \comm{\bm{a}}{\bm{b}}_{\mathrm{cl}}
  =\sum_i\qty(a_i\pdv{\bm{b}}{\phi_i}-b_i\pdv{\bm{a}}{\phi_i}).
  \label{eq:classical-lie-bracket}
\end{align}
With this notation, the lower-order HF expansion formulas of the deterministic force are given by~\cite{sato2025floquet}
\begin{align}
  \begin{aligned}
    \bm{f}_{\mathrm{eff}}
    &=\bm{f}_0+\bm{f}_{\mathrm{eff}}^{(1)}+\bm{f}_{\mathrm{eff}}^{(2)}
    +\mathcal{O}(\omega^{-3}),\\
    \bm{f}_{\mathrm{eff}}^{(1)}
    &=-i\sum_{m\neq0}\frac{\comm{\bm{f}_{-m}}{\bm{f}_m}_{\mathrm{cl}}}{2m\omega},\\
    \bm{f}_{\mathrm{eff}}^{(2)}
    &=-\sum_{m\neq0}\left\{
      \frac{\comm{\bm{f}_{-m}}{\comm{\bm{f}_0}{\bm{f}_m}_{\mathrm{cl}}}_{\mathrm{cl}}}{2m^2\omega^2}
      +\sum_{\substack{p\neq0\\p\neq m}}
      \frac{\comm{\bm{f}_{-p}}{\comm{\bm{f}_{p-m}}{\bm{f}_m}_{\mathrm{cl}}}_{\mathrm{cl}}}{3mp\omega^2}
    \right\}.
  \end{aligned}
  \label{eq:deterministic-HF-expansion}
\end{align}
For the single-color CPL drive, only the Fourier components $\bm f_{m}$ with $m=0,\pm1$ are nonzero.
In this case, the third-order term $\bm{f}_{\mathrm{eff}}^{(3)}$ is given by
\begin{align}
  \bm{f}_{\mathrm{eff}}^{(3)} = i\sum_{m=\pm1} \qty{\frac{\qty[\bm{f}_{-m},\qty[\bm{f}_{0}, \qty[\bm{f}_{0},\bm{f}_{m}]_{\mathrm{cl}}]_{\mathrm{cl}}]_{\mathrm{cl}}}{2m^3\omega^{3}} + \sum_{p\neq0} \frac{\qty[\bm{f}_{-p},\qty[\bm{f}_{p}, \qty[\bm{f}_{-m},\bm{f}_{m}]_{\mathrm{cl}}]_{\mathrm{cl}}]_{\mathrm{cl}}}{12mp^{2}\omega^{3}}}.
  \label{eq:ULE-Hf-expansion}
\end{align}

As shown below, for the multipolar model, the OLE yields a nonzero rotational force already at order $\omega^{-1}$ from Eq.~\eqref{eq:deterministic-HF-expansion} because $\qty[\bm{f}_{-1},\bm{f}_1]_{\mathrm{cl}} \neq \bm 0$.
In the ULE, the rotational contributions through order $\omega^{-2}$ vanish because $\qty[\bm{f}_{-1},\bm{f}_1]_{\mathrm{cl}} = \qty[\bm{f}_{-1},\qty[\bm{f}_0,\bm{f}_1]_{\mathrm{cl}}]_{\mathrm{cl}} = \bm 0$.
At the next order, $\qty[\bm{f}_{-1},\qty[\bm{f}_{0},\qty[\bm{f}_0,\bm{f}_1]_{\mathrm{cl}}]_{\mathrm{cl}}]_{\mathrm{cl}} \neq \bm 0$ yields the first nonzero rotational force, which is of order $\omega^{-3}$.
The second term in Eq.~\eqref{eq:ULE-Hf-expansion} vanishes because $\qty[\bm{f}_{-1},\bm{f}_{1}]_{\mathrm{cl}}=\bm{0}$.
The model-specific calculations below give the explicit form of the leading rotational forces and show that the next sub-leading corrections occur at orders $\omega^{-3}$ for the OLE and $\omega^{-5}$ for the ULE.

%%%%%%%%%%%%%%%%%%%%%%%%%%%%%%%%%%
\subsection{High-frequency Expansion for the OLE of the Electric Dipolar Model}\label{sec:OLE-HF-expansion}
Here, we explain the logical flow of deriving the effective EOM for the OLE of the dipolar ($n=1$) model.
We introduce the ``center-of-mass'' coordinate $\bm{R}=(\bm{r}_{1}+\bm{r}_{2})/2=(X,Y)$ and the relative coordinate $\bm{r}=\bm{r}_{2}-\bm{r}_{1}=(\ell\cos\theta,\ell\sin\theta)$.
Rewriting the EOM for the classical dynamical variables $\bm{\phi}(t)=[X(t),Y(t),\theta(t)]^{\top}$ in the form of the stochastic differential equation (SDE), $\dot{\bm{\phi}} = \bm{f}(\bm{\phi},t) + G(\bm{\phi})\bm{h}$, we obtain
\begin{align}
  \bm{f}(\bm{\phi},t) &= %\coloneq
  \left[ -\frac{\kappa}{\gamma}X,
    -\frac{\kappa}{\gamma}Y,
  -\frac{2qE_0}{\gamma \ell}\sin(\omega t - \theta)\right]^\top,\\
  G & =  %\coloneq
  \qty{g_{ij}}_{1\leq i\leq3,1\leq j\leq 4},\\
  \bm{h} &=  %\coloneq
  [h^{(1)}_{x}(t), h^{(1)}_{y}(t), h^{(2)}_{x}(t), h^{(2)}_{y}(t)]^{\top}.
\end{align}
In the matrix $G$, $g_{11}=g_{13}=g_{22}=g_{24}=1/(2\gamma)$,
$g_{31}=-g_{33}=\sin\theta/(\gamma \ell)$,
$g_{32}=-g_{34}=-\cos\theta/(\gamma \ell)$, and all other elements are zero.

Using the Floquet theory for classical nonlinear systems in Refs.~\cite{higashikawa2018floquet,sato2025floquet}, we perform the HF expansion for the FP equation corresponding to SDE (See Sec.~\ref{sec:HF-expansion-general}). As a result, we obtain the effective OLE in the form $\dot{\bm{\phi}} = \bm{f}^{\mathrm{eff}}(\bm{\phi}) + G^{\mathrm{eff}}(\bm{\phi})\bm{h}$.
In the leading order, we have
\begin{align}
  \bm{f}^{\mathrm{eff}} &= \left[-\frac{\kappa}{\gamma}X, -\frac{\kappa}{\gamma}Y, \Omega_{1}^{\mathrm{O}}\right]^{\top}+ \mathcal{O}(\omega^{-3}),\\
  G^{\mathrm{eff}} & = G + \mathcal{O}(\omega^{-3})
\end{align}
Focusing only on the rotational motion, we find that when the thermal-noise contribution can be neglected at $T=0$, the steady state rotates with the angular frequency $\dot{\theta}\simeq\Omega_{1}^{\mathrm{O}} + \mathcal{O}(\omega^{-3})$.

%%%%%%%%%%%%%%%%%%%%%%%%%%%%%%%%%%
\subsection{HF Expansion for the ULE of the Electric Dipolar Model: Spin and Orbit}
Next, we lead to the HF-expanded effective ULE for the dipolar model.
For Eq.~(1) with $n=1$, we define the total mass $M= m_{1}+m_{2}$, the reduced mass $\mu_{1}= m_{1}m_{2}/M$, the center-of-mass (COM) coordinate $\bm{R}=(m_{1}\bm{r}_{1}+m_{2}\bm{r}_{2})/M=(X,Y)$ and the relative coordinate $\bm{r}$.
For simplicity, we use the same symbol $\bm{R}$ as that in Sec.~\ref{sec:OLE-HF-expansion}, although its meaning is slightly different from that in the OLE case; in both cases, $\bm{R}$ represents the position of the Floquet rotation axis.
Let us introduce the velocities $v_{X}$, $v_{Y}$, and $v_{\theta}$ ($v_a=da/dt$), and then define the vector of the classical dynamical variables $\bm{\phi}(t)=[X(t),v_{X}(t),Y(t),v_{Y}(t),\theta(t),v_{\theta}(t)]^{\top}$ in the SDE $\dot{\bm{\phi}} = \bm{f}(\bm{\phi},t) + G(\bm{\phi})\bm{h}$.
In this notation, the vector
$
\bm{f}(\bm{\phi},t) =
\qty[
  v_{X},
  f_{X}\qty(\bm{\phi}),
  v_{Y},
  f_{Y}\qty(\bm{\phi}),
  v_{\theta},
  f_{\theta}\qty(\bm{\phi}) + f_{\mathrm{ext}}(\theta,t)
]^\top
$ is given by
\begin{align}
  &f_{X}=-\frac{\gamma}{M}\qty[2v_{X}\! -\eta_m\ell v_{\theta}\sin\theta] -\frac{\kappa}{M}\qty[2X\! + \eta_m\ell\cos\theta],\label{eq:ULE-dipole-6variables-fX}\\
  &f_{Y}=-\frac{\gamma}{M}\qty[2v_{Y}\! + \eta_m\ell v_{\theta}\cos\theta] -\frac{\kappa}{M}\qty[2Y\! +\eta_m\ell\sin\theta],\label{eq:ULE-dipole-6variables-fY}\\
  &f_{\theta}
  =-\nu_{m}\gamma v_{\theta} + \frac{\eta_m\gamma}{\mu_1\ell}\qty[\qty(v_{X} + \frac{\kappa \ell}{\gamma} X)\sin\theta
  - \qty(v_{Y} + \frac{\kappa \ell}{\gamma} Y)\cos\theta],\\
  &f_{\mathrm{ext}}(\theta,t) = - \frac{qE_{0}}{\mu_{1} \ell}\sin\qty(\omega t - \theta),\label{eq:ULE-dipole-6variables-ftheta}
\end{align}
where $\nu_{m}=\mu_1^{-1}-2M^{-1}$,
$\Delta m=m_1-m_2$, and $\eta_m = \Delta m/M$.
The non-zero matrix elements of $G=\{g_{ij}\}_{1\leq i\leq6,1\leq j\leq 4}$ are $g_{21}=g_{23}=g_{42}=g_{44}=1/M$, $g_{61}=\sin\theta/(m_{1}\ell)$, $g_{62}=-\cos\theta/(m_{1}\ell)$, $g_{63}=-\sin\theta/(m_{2}\ell)$, and $g_{64}=\cos\theta/(m_{2}\ell)$.

Applying the HF expansion method in Refs.~\cite{higashikawa2018floquet,sato2025floquet}, we obtain the effective ULE $\dot{\bm{\phi}}=\bm{f}^{\mathrm{eff}}(\bm{\phi})+G^{\mathrm{eff}}(\bm{\phi})\bm{h}$ with
\begin{align}
  \bm{f}^{\mathrm{eff}}(\bm{\phi}) &=
  \left[
    v_{X},
    f_{X}\qty(\bm{\phi}'),
    v_{Y},
    f_{Y}\qty(\bm{\phi}'),
    v_{\theta} + \Omega_{1}^{\mathrm{U}},
    f_{\theta}\qty(\bm{\phi})
  \right]^\top
  + \mathcal{O}(\omega^{-5}),\\
  G^{\mathrm{eff}}(\bm{\phi}) &= G(\bm{\phi}) + \mathcal{O}(\omega^{-5}),
\end{align}
where $\bm{\phi}'=\left(X,v_{X},Y,v_{Y},\theta,v_{\theta} + \Omega_{1}^{\mathrm{U}}\right)^{\top}$.
As a consequence, for $m_{1}=m_{2}$, the steady state rotation frequency is given by $\dot{\theta}\simeq\Omega_{1}^{\mathrm{U}}+\mathcal{O}(\omega^{-5})$.

In the mass-imbalance case of $m_{1}\neq m_{2}$,
we have to carefully define two kinds of CPL-driven rotations: autorotation (spin) and revolution (orbital motion). 
The autorotation (spin motion) stands for the rotation that we have discussed so far (see e.g., $\dot{\theta}\simeq\Omega_{1}^{\mathrm{U}}+\mathcal{O}(\omega^{-5})$), while the revolution (orbital motion) is the rotation of the COM coordinate $\bm R=(X,Y)$. In fact, Eqs.~\eqref{eq:ULE-dipole-6variables-fX} and \eqref{eq:ULE-dipole-6variables-fY} show that $v_\theta$ does not affect the force $(f_X,f_Y)$ for $m_1=m_2$ ($\Delta m=0$), whereas it does for $m_1\neq m_2$. As a result, the orbital motion driven by CPL can occur only in the mass-imbalance case.

To discuss the steady-state motion for $m_{1}\neq m_{2}$, we consider a steady solution of the effective ULE in a rotating frame that undergoes uniform rotation with angular velocity $\Omega_{\mathrm{rot}}$.
We introduce transformed variables in $\Omega_{\mathrm{rot}}$-rotating-frame coordinates by $\bar{\bm{\phi}}=\left(\bar{X},\bar{v}_{X},\bar{Y},\bar{v}_{Y},\bar{\theta},\bar{v}_{\theta}\right)^{\top}$,
and we impose the steady-state condition $\dot{\bar{\bm{\phi}}} = \bm{0}$.
Then we obtain the quintic equation for $\Omega_{\mathrm{rot}}$,
\begin{align}
  \Omega_{\mathrm{rot}}^{5} - \Omega_{1}^{\mathrm{U}}\Omega_{\mathrm{rot}}^{4} &+ (1-\tilde{M})K_{1}\Omega_{\mathrm{rot}}^{3} - K_{1}\Omega_{1}^{\mathrm{U}}\Omega_{\mathrm{rot}}^{2} + (1-\tilde{M})K_{2}\Omega_{\mathrm{rot}} - K_{2}\Omega_{1}^{\mathrm{U}} = 0. \label{eq:dipole-Omega-equation}
\end{align}
Here, $\tilde{M},K_{1},K_{2}$ are defined by $\tilde{M}= (1/2)(m_{1}-m_{2})^{2}/(m_{1}^{2}+m_{2}^{2})$,
$K_{1} = 4(\gamma^2/M^{2}-\kappa/M)$, and $K_{2} = 4\kappa^{2}/M^{2}$.
Note that $\tilde{M}=0$ for $m_{1}=m_{2}$, while $\tilde{M}=1/2$ in the limit of $(m_{1},m_{2})\to (M,0)$ or $(0,M)$. We note that the assumption of $\dot{\bar{\bm{\phi}}} = \bm{0}$ means that both spin and orbital rotations have the same rotation frequency $\Omega_{\rm rot}$.

Although it is difficult to solve Eq.~\eqref{eq:dipole-Omega-equation} exactly, in the HF regime we assume $\Omega_{\mathrm{rot}}\simeq\Omega_{1}^{\mathrm{U}}\qty(1+\delta + \mathcal{O}\qty(\delta^{2}))$ ($|\delta|\ll1$).
Then, we obtain
\begin{align}
  &\delta = \frac{\tilde{M}\qty(K_{1}\qty(\Omega_{1}^{\mathrm{U}})^{2}+K_{2})}{\qty(\Omega_{1}^{\mathrm{U}})^{4}+(1-3\tilde{M})K_{1}\qty(\Omega_{1}^{\mathrm{U}})^{2}+(1-\tilde{M})K_{2}}.
  \label{eq:Delta}
\end{align}
Consequently, for the mass-imbalance case $m_1\neq m_2$, if both $\Omega_{1}^{\mathrm{U}}/\sqrt{\kappa/M}$ and $\tilde M$ are sufficiently small, the steady-state rotation frequency is indeed given by $\dot{\theta}\simeq\Omega_{1}^{\mathrm{U}}\qty(1+\delta+\mathcal{O}(\delta^2))+\mathcal{O}(\omega^{-5})$.
As we mentioned in the main text, the results obtained by solving the ULE agree, for both spin and orbit, with the numerical solution of Eq.~\eqref{eq:dipole-Omega-equation}, and $\Omega_{1}^{\mathrm{U}}(1+\delta)$ from Eq.~\eqref{eq:Delta} also provides a good approximation. The fact of $\delta>0$ means that a mass imbalance causes the rotation frequency to increase.

%%%%%%%%%%%%%%%%%%%%%%%%%%%%%%%
\subsection{HF Expansion in Underdamped Multipolar Models}\label{Sec:Supplemental-HF-under-multipole}
In this section, we present the derivation of the Floquet rotation for the electric multipolar model, which is summarized in Table~\ref{tab:frequency-table}.
We first derive the effective ULE via the HF expansion for the electric $2^{n}$-polar model described by Eq.~(1) in the main text.
We introduce the total mass $M=\sum_{j=1}^{2n}m_{j}$, the COM coordinate $\bm{R}=\sum_{j=1}^{2n}m_{j}\bm{r}_{j}/M\qty(=\qty(X,Y))$, and the geometric center $\bm{\tilde{R}}=\sum_{j=1}^{2n}\bm{r}_{j}/(2n)$.
We define $\bm{r}=\bm{R}-\bm{\tilde{R}}=d\qty(\cos\qty(\theta+\theta_{0}),\sin\qty(\theta+\theta_{0}))$, where $\theta_{0}$ is the angle between $\bm{r}$ and $\bm{r}_{n+1}-\bm{r}_{1}=\ell\qty(\cos\theta,\sin\theta)$.
The length of $d=|\bm r|$ is an important parameter for classifying the ULE of the multipolar model.
%%%%%%%%%%%%%%%%%%%%%%%%%%%%%%%%%%%%%
\begin{figure}[b]
  \includegraphics[width=\textwidth/2]{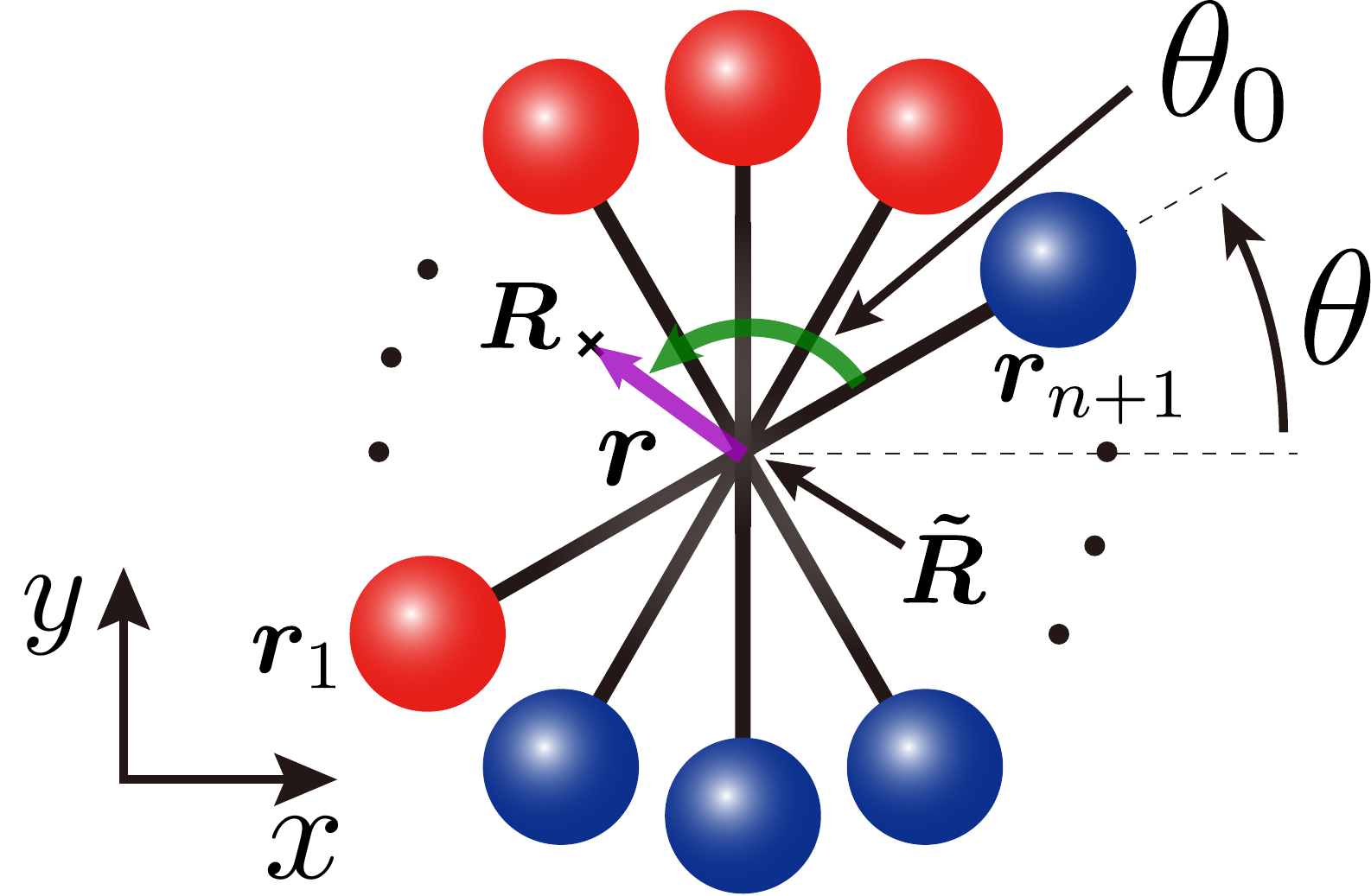}
  \caption{
    Schematic of the multipolar model.
    Point charges with masses $m_{j}$ ($j=1,\dots,2n$) are equally spaced on a circle of radius $\ell/2$.
    With the multipole COM coordinate $\bm{R}$, the geometric center $\bm{\tilde{R}}$, and the rotation angle $\theta$, we define $d=|\bm{R}-\bm{\tilde{R}}|$.
    The angle $\theta_{0}$ is the angle between $\bm{r}$ and $\bm{r}_{n+1}-\bm{r}_{1}=\qty(\ell\cos\theta,\ell\sin\theta)$.
  }
  \label{fig:multipoleCoordinate}
\end{figure}
%%%%%%%%%%%%%%%%%%%%%%%%%%%%%%%%%%%%%
With these definitions, the EOMs for the COM coordinate $\bm{R}$ and the rotation angle $\theta$ are given by
\begin{align}
  &\left\lbrace \,
  \begin{aligned}
    M\ddot{\bm{R}} =& -2\gamma\qty(\dot{\bm{R}} - \dot{\bm{r}}) - 2\kappa\qty(\bm{R} - \bm{r}) + \sum_{j=1}^{2n}\bm{h}^{(j)}\\
    \mu_{n}\ell^{2}\ddot{\theta} =& -2\gamma\qty{\qty(d^{2}+\frac{\ell^{2}}{4})\dot{\theta} + d\sin\qty(\theta + \theta_{0})\dot{X} - d\cos\qty(\theta+\theta_{0})\dot{Y}} \\
    & - \sum^{n}_{j=1}\frac{qE_{0}\ell}{n}\sin\qty(\omega t - \theta + \frac{\pi\qty(j-1)}{n})- 2\kappa d\qty{X\sin\qty(\theta+\theta_{0}) - Y\cos\qty(\theta+\theta_{0})} \\
    & + \sum^{2n}_{j=1}\left[h_{x}^{(j)}\qty{d\sin\qty(\theta+\theta_{0}) + \frac{\ell}{2}\sin\qty(\theta - \frac{\pi\qty(j-1)}{n})} \right.\\
    &\left. \qquad\qquad - h_{y}^{(j)}\qty{d\cos\qty(\theta+\theta_{0}) + \frac{\ell}{2}\cos\qty(\theta - \frac{\pi\qty(j-1)}{n})}\right].
  \end{aligned}
  \right.\label{eq:ULE-multipole-6variables-pre}
\end{align}
Here $\mu_{n}$ is defined as
\begin{align}
  \mu_{n}= M\qty(\frac{d^{2}}{\ell^{2}} + \frac{1}{4}) + \frac{d}{\ell}\sum^{2n}_{j=1}m_{j}\cos\qty(\theta_{0} + \frac{\pi\qty(j-1)}{n})\label{eq:mu_n}.
\end{align}
Introducing $v_{X},v_{Y},v_{\theta}$ and rewriting Eq.~\eqref{eq:ULE-multipole-6variables-pre} in the form of a SDE for the classical variables $\bm{\phi}(t)=\qty[X(t),v_{X}(t),Y(t),v_{Y}(t),\theta(t),v_{\theta}(t)]^{\top}$, $\dot{\bm{\phi}} = \bm{f}(\bm{\phi},t) + G(\bm{\phi})\bm{h}$, we obtain
\begin{align}
  &\bm{f}\qty(\bm{\phi},t) =
  \begin{pmatrix}
    v_{X}\\
    f_{X}\qty[X,v_{X},\theta,v_{\theta}] \\
    v_{Y}\\
    f_{Y}\qty[Y,v_{Y},\theta,v_{\theta}] \\
    v_{\theta}\\
    f_{\theta}\qty[X,v_{X},Y,v_{Y},\theta,v_{\theta}] - B_{n}^{\mathrm{U}}\sin\qty(\omega t - \theta + \frac{\pi\qty(n-1)}{2n})
  \end{pmatrix},
  \label{eq:ULE-multipole-6variables-f}\\
  &G\qty(\bm{\phi}) = \qty{g_{ij}}_{1\leq i\leq6,1\leq j\leq 4n}\ ,\quad
  \bm{h} =
  \begin{pmatrix}
    h^{(1)}_{x}(t)\\
    h^{(1)}_{y}(t)\\
    \vdots\\
    h^{(2n)}_{x}(t)\\
    h^{(2n)}_{y}(t)
  \end{pmatrix}.
  \label{eq:ULE-multipole-6variables-Gh}
\end{align}
The parameter $B_{n}^{\mathrm{U}}$ is given by $B_{n}^{\mathrm{U}}=\frac{qE_{0}}{n\mu_{n} \ell}\csc\qty(\frac{\pi}{2n})$ where $\csc x=1/\sin x$.
The components of $\bm f$ and $\bm h$ and the matrix elements of $G$ are defined as
\begin{align}
  &f_{X} = -\frac{2\gamma}{M}\qty{v_{X} + dv_{\theta}\sin\qty(\theta + \theta_{0})} - \frac{2\kappa}{M}\qty{X - d\cos\qty(\theta + \theta_{0})},\\
  &f_{Y} = -\frac{2\gamma}{M}\qty{v_{Y} - dv_{\theta}\cos\qty(\theta + \theta_{0})} - \frac{2\kappa}{M}\qty{Y - d\sin\qty(\theta + \theta_{0})},\\
  &f_{\theta} = -\frac{2\gamma}{\mu_{n} \ell^{2}}\qty{\qty(d^{2} + \frac{\ell^{2}}{4})v_{\theta} + d\sin\qty(\theta + \theta_{0})v_{X} - d\cos\qty(\theta + \theta_{0})v_{Y}}-\frac{2\kappa d}{\mu_{n}\ell^{2}}\qty{X\sin\qty(\theta+\theta_{0}) - Y\cos\qty(\theta + \theta_{0})},\\
  &g_{2i} =
  \begin{cases}
    0 & (\mathrm{where\ }i\ \mathrm{is\ even}) \\
    \dfrac{1}{M} & (\mathrm{where\ }i\ \mathrm{is\ odd})
  \end{cases},\qquad
  g_{4i} =
  \begin{cases}
    0 & (\mathrm{where\ }i\ \mathrm{is\ odd}) \\
    \dfrac{1}{M} & (\mathrm{where\ }i\ \mathrm{is\ even})
  \end{cases},\\
  &g_{6i} =
  \begin{cases}
    \qty{d\sin\qty(\theta+\theta_{0}) + \dfrac{\ell}{2}\sin\qty(\theta - \dfrac{\pi\qty(k-1)}{n})}\qty(\mu_{n}l^{2})^{-1} & (\mathrm{where\ }i=2k-1\ (k\in\mathbb{N})) \\
    -\qty{d\cos\qty(\theta+\theta_{0}) + \dfrac{\ell}{2}\cos\qty(\theta - \dfrac{\pi\qty(k-1)}{n})}\qty(\mu_{n}l^{2})^{-1} & (\mathrm{where\ }i=2k\ (k\in\mathbb{N}))
  \end{cases}.
\end{align}
All other $g_{ij}$ are zero.
We note that when $d=0$, that is, when the COM coordinate $\bm R$ coincides with the geometric center $\tilde{\bm R}$, $\bm R$ and $\theta$ become independent each other in the deterministic force matrix $\bm f$ (see the matrix elements of $\bm f$).

Following the classical Floquet theory in Refs.~\cite{sato2025floquet, higashikawa2018floquet} (See Sec.~\ref{sec:HF-expansion-general}), we can perform the HF expansion and express the effective EOM as
$\dot{\bm{\phi}}= \bm{f}^{\mathrm{eff}}(\bm{\phi}) + G^{\mathrm{eff}}(\bm{\phi})\bm{h}$.
Then we obtain
\begin{align}
  \label{eq:f_eff_MultipolarUML}
  &\bm{f}^{\mathrm{eff}}\qty(\bm{\phi}) =
  \begin{pmatrix}
    v_{X}\\
    f_{X}\qty[X,v_{X},\theta,v_{\theta} + \Omega_{n}^{\mathrm{U}}] \\
    v_{Y}\\
    f_{Y}\qty[Y,v_{Y},\theta,v_{\theta} + \Omega_{n}^{\mathrm{U}}] \\
    v_{\theta} + \Omega_{n}^{\mathrm{U}} \\
    f_{\theta}\qty[X,v_{X},Y,v_{Y},\theta,v_{\theta}]
  \end{pmatrix} + \mathcal{O}(\omega^{-5}),\\
  &\Omega_{n}^{\mathrm{U}}=\frac{A_{n}^{\mathrm{U}}}{\omega^3}\qty(\frac{qE_{0}}{\mu_{n} \ell}) ^{2},\\
  &G^{\mathrm{eff}}(\bm{\phi}) = G(\bm{\phi}) + \mathcal{O}(\omega^{-5})\label{eq:multipole-eff-ULE-6variables}.
\end{align}
Here we define the parameters $A_{n}^{\mathrm{U}}$ and $a_{n}$ as
\begin{align}
  A_{n}^{\mathrm{U}} &= \frac{a_{n}}{2n^{2}},\label{eq:A_n^ULE}\\
  a_{n} &= \csc^{2}\qty(\frac{\pi}{2n})\label{eq:a_n}.
\end{align}
For instance, $a_{2}=2$ and $a_{3}=4$.
In the limit $n\rightarrow\infty$, with $\mu_{n}\rightarrow \mu_{\infty} \qty(< Md/\ell)$, we have $\Omega_{n}^{\mathrm{U}}\rightarrow(2/\pi^{2})\qty(qE_{0}/(\mu_{\infty}\ell))^{2}\omega^{-3}$.
These results are summarized in Table~\ref{tab:frequency-table}.
The result of Eq.~\eqref{eq:f_eff_MultipolarUML} implies that for $d=0$, the steady state is given by the Floquet rotation with the frequency $\dot{\theta}\simeq\Omega_{n}^{\mathrm{U}}+\mathcal{O}(\omega^{-5})$.

To discuss the steady-state motion for $d\neq0$, in the same manner as for the dipole case, we consider a steady-state solution of the effective ULE in a rotating frame that undergoes uniform rotation with angular velocity $\Omega_{\mathrm{rot}}$.
We define the rotating-frame variables $\bar{\theta}$ and $\bar{v}_{\theta}$ as
\begin{align}
  \left\lbrace \,
  \begin{aligned}
    \bar{\theta} &= \theta - \Omega_{\mathrm{rot}} t\\
    \bar{v}_{\theta} &= v_{\theta} + \Omega_{n}^{\mathrm{U}} - \Omega_{\mathrm{rot}}
  \end{aligned}
  .\right.
\end{align}
The COM coordinate is transformed as
\begin{align}
  \begin{pmatrix}
    \bar{X}\\
    \bar{Y}
  \end{pmatrix}
  &=
  \begin{pmatrix}
    \cos(\Omega_{\mathrm{rot}} t) & \sin(\Omega_{\mathrm{rot}} t) \\
    -\sin(\Omega_{\mathrm{rot}} t) & \cos(\Omega_{\mathrm{rot}} t)
  \end{pmatrix}
  \begin{pmatrix}
    X\\
    Y
  \end{pmatrix}
\end{align}
and we define $\bar{v}_{X}=\dot{\bar{X}},\ \bar{v}_{Y}=\dot{\bar{Y}}$.
Letting $\bar{\bm{\phi}}=[\bar{X},\bar{v}_{X},\bar{Y},\bar{v}_{Y},\bar{\theta},\bar{v}_{\theta}]^{\top}$, we impose the steady-state condition $\dot{\bar{\bm{\phi}}} = \bm{0}$.
Then we obtain the quintic equation for $\Omega_{\mathrm{rot}}$,
\begin{align}
  \Omega_{\mathrm{rot}}^{5} - \Omega_{n}^{\mathrm{U}}\Omega_{\mathrm{rot}}^{4} + \frac{K_{1}}{4\tilde{\ell}+1}\Omega_{\mathrm{rot}}^{3} - K_{1}\Omega_{n}^{\mathrm{U}}\Omega_{\mathrm{rot}}^{2}  + \frac{K_{2}}{4\tilde{\ell}+1}\Omega_{\mathrm{rot}} - K_{2}\Omega_{n}^{\mathrm{U}} = 0 \label{eq:multipole-Omega-equation}
\end{align}
where $\tilde{\ell},K_{1},K_{2}$ are defined as $\tilde{\ell}= (d/\ell)^{2}$, $K_{1} = 4(\gamma^2/M^{2}-\kappa/M)$, and $K_{2} = 4\kappa^{2}/M^{2}$.
Although it is difficult to solve Eq.~\eqref{eq:multipole-Omega-equation} exactly, in the HF regime we assume
$\Omega_{\mathrm{rot}}\simeq\Omega_{n}^{\mathrm{U}}\qty(1+\delta_{n})$ with $|\delta_{n}|\ll1$.
Then, to first order in $\delta_{n}$, we obtain
\begin{align}
  &\Omega_{\mathrm{rot}} = \Omega_{n}^{\mathrm{U}}\qty(1 + \delta_{n} + \mathcal{O}\qty(\delta_{n}^{2}))\label{eq:multipole-Omega-delta(1)}\\
  &\delta_{n} =\frac{4\tilde{\ell}\qty(K_{1}\qty(\Omega_{n}^{\mathrm{U}})^{2}+K_{2})}{\qty(4\tilde{\ell}+1)\qty(\Omega_{n}^{\mathrm{U}})^{4}+\qty(1-8\tilde{\ell})K_{1}\qty(\Omega_{n}^{\mathrm{U}})^{2}+K_{2}}.
  \label{eq:multipole-Delta}
\end{align}

%%%%%%%%%%%%%%%%%%%%%%%%%%%%%%%
\subsection{HF Expansion in Overdamped Multipolar Models}\label{Sec:Supplemental-HF-over-multipole}
Similarly to the previous section, we here present the HF-expansion result for the electric $2^{n}$-pole described by the OLE, as in Eq.~(3) of the main text.
Applying the coordinate transformation to the COM coordinate
$\bm{R}=\sum_{j=1}^{2n}\bm{r}_{j}/(2n)\qty(=\qty(X,Y))$
and the relative coordinate
$\bm{r}=\bm{r}_{n+1}-\bm{r}_{1}=\ell\qty(\cos\theta,\sin\theta)$,
we obtain
\begin{align}
  &\left\lbrace \,
  \begin{aligned}
    2\gamma\dot{\bm{R}} &= - 2\kappa\bm{R} + \sum_{j=1}^{2n}\bm{h}^{(j)}\\
    \gamma \ell\dot{\theta} &= -\frac{2qE_{0}}{n}\sum_{j=1}^{n}\sin\qty(\omega t - \theta + \frac{\pi\qty(j-1)}{n}) \\
    &\quad + \sum_{j=1}^{2n} \left[h^{(j)}_{x}\sin\qty(\theta - \frac{\pi\qty(j-1)}{n}) - h^{(j)}_{y}\cos\qty(\theta - \frac{\pi\qty(j-1)}{n})\right].\\
  \end{aligned}
  \right.\label{eq:OLE-multipole-3variables-pre}
\end{align}
By rewriting Eq.~\eqref{eq:OLE-multipole-3variables-pre} in terms of $\bm{\phi}(t)=\qty[X(t),Y(t),\theta(t)]^{\top}$, we can express it in the form of an SDE $\dot{\bm{\phi}}=\bm{f}\qty(\bm{\phi},t) + G\qty(\bm{\phi})\bm{h}$, where
\begin{align}
  &\bm{f}\qty(\bm{\phi},t) \coloneq
  \begin{pmatrix}
    -\frac{\kappa}{\gamma}X \\
    -\frac{\kappa}{\gamma}Y \\
    -B_{n}^{\mathrm{O}}\sin\qty(\omega t - \theta + \frac{\pi\qty(n-1)}{2n})
  \end{pmatrix},\quad B_{n}^{\mathrm{O}}\coloneq\frac{2qE_{0}}{n\gamma \ell}\csc\qty(\frac{\pi}{2n}),\label{eq:OLE-multipole-3variables-f}\\
  & G\qty(\bm{\phi}) \coloneq \qty{g_{ij}}_{1\leq i\leq3,1\leq j\leq 2n}=
  \begin{pmatrix}
    \frac{1}{2\gamma} & 0 & \cdots & \frac{1}{2\gamma} & 0 \\
    0 & \frac{1}{2\gamma} & \cdots & 0 & \frac{1}{2\gamma} \\
    \frac{\sin\theta}{\gamma \ell} & -\frac{\cos\theta}{\gamma \ell} & \cdots & \frac{\sin\qty(\theta + \pi/n)}{\gamma \ell} & -\frac{\cos\qty(\theta + \pi/n)}{\gamma \ell}
  \end{pmatrix}
  ,\quad \bm{h} \coloneq
  \begin{pmatrix}
    h^{(1)}_{x}(t)\\
    h^{(1)}_{y}(t)\\
    \vdots\\
    h^{(2n)}_{x}(t)\\
    h^{(2n)}_{y}(t)
  \end{pmatrix}.\label{eq:OLE-multipole-3variables}
\end{align}
Performing the HF expansion and expressing the effective EOM in the form
$\dot{\bm{\phi}} = \bm{f}^{\mathrm{eff}}(\bm{\phi}) + G^{\mathrm{eff}}(\bm{\phi})\bm{h}$,
we obtain
\begin{align}
  &\bm{f}^{\mathrm{eff}}\qty(\bm{\phi}) =
  \begin{pmatrix}
    -\frac{\kappa}{\gamma}X \\
    -\frac{\kappa}{\gamma}Y \\
    \Omega_{n}^{\mathrm{O}}
  \end{pmatrix} + \mathcal{O}(\omega^{-3}),\notag\\
  &\Omega_{n}^{\mathrm{O}}\coloneq\frac{A_{n}^{\mathrm{O}}}{\omega}\qty(\frac{qE_{0}}{\gamma \ell})^{2},\notag\\
  &G^{\mathrm{eff}}(\bm{\phi}) = G(\bm{\phi}) + \mathcal{O}(\omega^{-3}).\label{eq:multipole-eff-OLE-3variables}
\end{align}
Here
\begin{align}
  A_{n}^{\mathrm{O}} &\coloneq \frac{2a_{n}}{n^{2}}\label{eq:A_n^OLE}
\end{align}
and in the limit $n\rightarrow\infty$ we find
$\Omega_{n}^{\mathrm{O}} \to (8/\pi^{2})\qty(qE_{0}/(\gamma\ell))^{2}\omega^{-1}$.
These results show that, in the steady state, Floquet rotation has the rotation speed
$\dot{\theta}\simeq\Omega_{n}^{\mathrm{O}}+\mathcal{O}(\omega^{-3})$.

%%%%%%%%%%%%%%%%%%%%%%%%%%%%%%%%%%%%%%%%%%%%%%%%%%
%%%%%%%%%%%%%%%%%%%%%%%%%%%%%%%%%%%%%%%%%%%%%%%%%%
\subsection{Floquet Rotation in Multipolar Models}\label{Sec:Supplemental-HF-multipole}
From the results in Secs.~\ref{sec:OLE-HF-expansion}-\ref{Sec:Supplemental-HF-over-multipole}, we have understood the Floquet rotation in the multipolar models driven by CPL, especially, in the case of equal masses $m_1=m_2=\cdots=m_{2n}$. The Floquet rotation frequencies in the HF regime are summarized in Table~\ref{tab:frequency-table}.
When we move to the dipole, quadrupole, and large-$n$ limits, the $\omega$ dependence of the Floquet rotation frequencies in the OLE and ULE remain $\omega^{-1}$ and $\omega^{-3}$ respectively, while the prefactors of the frequencies depend on the multipolar parameter $n$.

%%%%%%%%%%%%%%%%%%%%%%%%%%%%%%%%%%%%%%%
\renewcommand{\arraystretch}{1.15}
\begin{table}[tb]
  \caption{
    Prefactors of Floquet rotation from the HF expansion.
    The upper [lower] row shows the coefficient $A_{n}^{\mathrm{O}}$ [$A_{n}^{\mathrm{U}}$] in $\Omega_{n}^{\mathrm{O}}=A_{n}^{\mathrm{O}}(qE_{0}/(\gamma \ell))^{2}\omega^{-1}$ [$\Omega_{n}^{\mathrm{U}}=A_{n}^{\mathrm{U}}(qE_{0}/(\mu_{n}\ell))^{2}\omega^{-3}$], derived from the effective OLE [effective ULE].
    Here $a_{n} = \csc^{2}\qty(\frac{\pi}{2n})$, with $a_{n}/n^{2}\to 4/\pi^{2}$ as $n\to\infty$.
    The explicit form of $\mu_{n}$ is given in Eq.~\eqref{eq:mu_n}.
  }
  \label{tab:frequency-table}
  \begin{ruledtabular}
    \begin{tabular}{cccccc}
      & $n=1$ & $n=2$ & $n=3$ & $n$ & $n\to\infty$ \\
      \hline
      $A_{n}^{\mathrm{O}}$ & $2$ & $1$ & $8/9$ & $2a_{n}/n^{2}$ & $8/\pi^{2}$ \\
      $A_{n}^{\mathrm{U}}$ & $1/2$ & $1/4$ & $2/9$ & $a_{n}/(2n^{2})$ & $2/\pi^{2}$ \\
    \end{tabular}
  \end{ruledtabular}
\end{table}
%%%%%%%%%%%%%%%%%%%%%%%%%%%%%%%%%%%%%%%

%%%%%%%%%%%%%%%%%%%%%%%%%%%%%%%%%%%%%%%%%%%%%%
\subsection{Mode-Separation Method}
In this section, as an alternative route to derive effective EOMs, we introduce the method of mode separation.

The mode-separation method~\cite{Kapitza1951ZhETF,landaulifshitz1976mechanics,zhang2025spin} decomposes dynamical variables in the EOMs into an HF component (small oscillations arising from the linear response to the external drive) and a low-frequency component (slow, averaged motion). By averaging out the HF component, one can extract an effective EOM governing the slow dynamics on long time scales.
Introducing a slow variable $\theta_{\mathrm{slow}}(t)$ and a fast variable $\theta_{\mathrm{fast}}(t)$, we decompose
\begin{align}
  \theta &= \theta_{\mathrm{slow}} + \theta_{\mathrm{fast}}.\label{eq:multipole-mode-separation}
\end{align}
We also define the time average over $\mathcal{T}=2\pi/\omega$ by $\langle\ \cdots\ \rangle_{\mathcal{T}} \coloneq \frac{1}{\mathcal{T}}\int_{t}^{t+\mathcal{T}} \cdots\ dt'$.
Assuming that the fast motion is periodic in time, $\theta_{\mathrm{fast}}(t)=\theta_{\mathrm{fast}}(t+\mathcal{T})$, and has zero mean, $\langle \theta_{\mathrm{fast}}\rangle_{\mathcal{T}}=0$, we find
\begin{align}
  \langle \theta \rangle_{\mathcal{T}} &= \langle\theta_{\mathrm{slow}}\rangle_{\mathcal{T}} + \langle \theta_{\mathrm{fast}} \rangle_{\mathcal{T}}
  =\langle\theta_{\mathrm{slow}}\rangle_{\mathcal{T}}.
  \label{eq:time-average-theta}
\end{align}
Furthermore, since $\theta_{\mathrm{slow}}$ varies much more slowly than $\theta_{\mathrm{fast}}$, we approximate
\begin{align}
  \langle \theta_{\mathrm{slow}} \rangle_{\mathcal{T}} &\simeq \theta_{\mathrm{slow}}.\label{eq:slow-variable-approx}
\end{align}

First, we focus on the OLE of the multipolar model [the third component of Eq.~\eqref{eq:OLE-multipole-3variables-f}], neglecting the noise term (i.e., at $T=0$). In this case, (as we already mentioned) $\bm R$ and $\theta$ are decomposed in the EOM, and we have a closed differential equation for $\theta$:
\begin{align}
  \dot{\theta} &= -B_{n}^{\mathrm{O}}\sin\qty(\omega t - \theta + \frac{\pi\qty(n-1)}{2n}).\label{eq:OLE-multipole-3variables-theta-f}
\end{align}
Substituting Eq.~\eqref{eq:multipole-mode-separation}, we obtain
\begin{align}
  &\dot{\theta}_{\mathrm{slow}} + \dot{\theta}_{\mathrm{fast}} = -B_{n}^{\mathrm{O}}\sin\qty(\omega t - \theta_{\mathrm{fast}} - \tilde{\theta}_{\mathrm{slow}}) \label{eq:OLE-multipole-theta-separation},\\
  &\tilde{\theta}_{\mathrm{slow}} =\theta_{\mathrm{slow}} - \frac{\pi(n-1)}{2n}.
\end{align}
As the HF condition under which mode separation holds, we assume
\begin{align}
  \epsilon_{\mathrm{O}}= \frac{B_{n}^{\mathrm{O}}}{\omega}\ll1\qquad\qty(\Leftrightarrow\ \omega\gg B_{n}^{\mathrm{O}}). \label{eq:epsilon_def}
\end{align}
Moreover, we assume that $\theta_{\mathrm{fast}}=\mathcal{O}(\epsilon_{\mathrm{O}})$ and  $|\theta_{\mathrm{fast}}|\ll1$ hold (as verified below).
Expanding the right-hand side of Eq.~\eqref{eq:OLE-multipole-theta-separation} up to second order in $\theta_{\mathrm{fast}}$, we obtain
\begin{align}
  \dot{\theta}_{\mathrm{slow}}+\dot{\theta}_{\mathrm{fast}}
  =-B_{n}^{\mathrm{O}}\qty[\sin\qty(\omega t - \tilde{\theta}_{\mathrm{slow}})-\theta_{\mathrm{fast}}\cos\qty(\omega t - \tilde{\theta}_{\mathrm{slow}}) - \frac{\theta_{\mathrm{fast}}^{2}}{2}\sin\qty(\omega t - \tilde{\theta}_{\mathrm{slow}})] + \mathcal{O}\qty(\epsilon_{\mathrm{O}}^{4}\omega). \label{eq:OLE-multipole-theta-separation-2nd-expansion}
\end{align}
Taking the time average of Eq.~\eqref{eq:OLE-multipole-theta-separation-2nd-expansion}, and using
$\langle\dot{\theta}_{\mathrm{fast}}\rangle_{\mathcal{T}}=0$ and $\left\langle\sin(\omega t - \tilde{\theta}_{\mathrm{slow}}) \right\rangle_{\mathcal{T}} = 0$, we get
\begin{align}
  \dot{\theta}_{\mathrm{slow}} = B_{n}^{\mathrm{O}}\left\langle \theta_{\mathrm{fast}}\cos\qty(\omega t - \tilde{\theta}_{\mathrm{slow}})\right\rangle_{\mathcal{T}} + \frac{B_{n}^{\mathrm{O}}}{2}\left\langle \theta_{\mathrm{fast}}^{2}\sin\qty(\omega t - \tilde{\theta}_{\mathrm{slow}})\right\rangle_{\mathcal{T}} + \mathcal{O}\qty(\epsilon_{\mathrm{O}}^{4}\omega). \label{eq:slow_eq_with_quadratic}
\end{align}

In what follows, we perform a perturbative expansion $\theta_{\mathrm{fast}}=\theta_{\mathrm{fast}}^{(1)}+\theta_{\mathrm{fast}}^{(2)}+\mathcal{O}(\epsilon_{\mathrm{O}}^{3})$, where the superscript $(k)$ denotes the order in $\epsilon_{\mathrm{O}}$,
i.e., $\theta_{\mathrm{fast}}^{(k)}=\mathcal{O}(\epsilon_{\mathrm{O}}^{k})$.
From Eq.~\eqref{eq:slow_eq_with_quadratic} we have $\dot{\theta}_{\mathrm{slow}}=\mathcal{O}(\epsilon_{\mathrm{O}}^{2}\omega)$, so Eq.~\eqref{eq:OLE-multipole-theta-separation-2nd-expansion} can be written as
\begin{align}
  \dot{\theta}_{\mathrm{fast}} &= - B_{n}^{\mathrm{O}}\sin\qty(\omega t - \tilde{\theta}_{\mathrm{slow}}) + \mathcal{O}\qty(\epsilon_{\mathrm{O}}^{2}\omega)\label{eq:theta_f-1st}.
\end{align}
On the time scale of $\theta_{\mathrm{fast}}$, $\theta_{\mathrm{slow}}$ can be regarded as approximately constant.
Integrating Eq.~\eqref{eq:theta_f-1st} with the integration constant set to zero, we obtain
\begin{align}
  \theta_{\mathrm{fast}} = \theta_{\mathrm{fast}}^{(1)} + \mathcal{O}\qty(\epsilon_{\mathrm{O}}^{2}),\quad \theta_{\mathrm{fast}}^{(1)} = \epsilon_{\mathrm{O}}\cos\qty(\omega t - \tilde{\theta}_{\mathrm{slow}})\label{eq:theta-fast-solution}
\end{align}
which confirms $\theta_{\mathrm{fast}}=\mathcal{O}(\epsilon_{\mathrm{O}})$.
Substituting Eq.~\eqref{eq:theta-fast-solution} into Eq.~\eqref{eq:slow_eq_with_quadratic}, we find
\begin{align}
  \dot{\theta}_{\mathrm{slow}} &= B_{n}^{\mathrm{O}}\epsilon_{\mathrm{O}}\left\langle\cos^{2}\qty(\omega t - \tilde{\theta}_{\mathrm{slow}})\right\rangle_{\mathcal{T}}  + \mathcal{O}\qty(\epsilon_{\mathrm{O}}^{3}\omega)
  = \frac{\qty(B_{n}^{\mathrm{O}})^{2}}{2\omega}  + \mathcal{O}\qty(\epsilon_{\mathrm{O}}^{3}\omega)\label{eq:theta_s-before-average}.
\end{align}

Next, we consider the second-order correction $\theta_{\mathrm{fast}}^{(2)}$.
Among the terms on the right-hand side of Eq.~\eqref{eq:OLE-multipole-theta-separation-2nd-expansion}, the contribution of order $\mathcal{O}(\epsilon_{\mathrm{O}}^{2}\omega)$ is $B_{n}^{\mathrm{O}}\theta_{\mathrm{fast}}^{(1)}\cos\qty(\omega t - \tilde{\theta}_{\mathrm{slow}})$, whose time average gives
\begin{align}
  \dot{\theta}_{\mathrm{slow}}^{(2)} &= B_{n}^{\mathrm{O}}\left\langle \theta_{\mathrm{fast}}^{(1)}\cos\qty(\omega t - \tilde{\theta}_{\mathrm{slow}})\right\rangle_{\mathcal{T}}
  = \frac{\qty(B_{n}^{\mathrm{O}})^{2}}{2\omega} \label{eq:theta-slow-(2)}.
\end{align}
Subtracting Eq.~\eqref{eq:theta-slow-(2)} from Eq.~\eqref{eq:OLE-multipole-theta-separation-2nd-expansion}, we obtain
\begin{align}
  \dot{\theta}_{\mathrm{fast}}^{(2)} &= B_{n}^{\mathrm{O}}\qty[\theta_{\mathrm{fast}}^{(1)}\cos\qty(\omega t - \tilde{\theta}_{\mathrm{slow}}) -   \frac{B_{n}^{\mathrm{O}}}{2\omega}]
  = \frac{\epsilon_{\mathrm{O}}^{2}\omega}{2}\cos\qty[2\qty(\omega t - \tilde{\theta}_{\mathrm{slow}})] \label{eq:theta_f-2nd}
\end{align}
and hence
\begin{align}
  \theta_{\mathrm{fast}}^{(2)} = \frac{\epsilon_{\mathrm{O}}^{2}}{4}\sin\qty[2\qty(\omega t - \tilde{\theta}_{\mathrm{slow}})] \label{eq:theta-fast-2nd-solution}.
\end{align}
Since $\left\langle \theta_{\mathrm{fast}}^{(2)}\cos\qty(\omega t - \tilde{\theta}_{\mathrm{slow}})\right\rangle_{\mathcal{T}}=0$ and $\left\langle (\theta_{\mathrm{fast}}^{(1)})^{2}\sin\qty(\omega t - \tilde{\theta}_{\mathrm{slow}})\right\rangle_{\mathcal{T}}=0$, the $\mathcal{O}(\epsilon_{\mathrm{O}}^{3}\omega)$ contribution in Eq.~\eqref{eq:theta_s-before-average} actually vanishes.
Therefore, returning to the original parameters, we obtain
\begin{align}
  \dot{\theta}_{\mathrm{slow}} &= \frac{\qty(B_{n}^{\mathrm{O}})^{2}}{2\omega}  + \mathcal{O}\qty(\epsilon_{\mathrm{O}}^{4}\omega)\\
  &=\frac{2}{\omega}\qty(\frac{qE_{0}}{n\gamma l})^{2}\csc^{2}\qty(\frac{\pi}{2n}) + \mathcal{O}\qty(\omega^{-3})\label{eq:theta_s-before-average-params}
\end{align}
This result is consistent with the effective OLE derived from the HF expansion in Floquet theory, $\dot{\theta}\simeq\Omega_{n}^{\mathrm{O}}+\mathcal{O}(\omega^{-3})$.

In contrast, for the ULE at $T=0$, the EOM does not take a closed form solely for $\theta$; rather, it is given as a set of coupled second-order differential equations for the three variables $X$, $Y$, and $\theta$.
Therefore, unlike the OLE case, it is difficult to apply mode separation directly to the $\theta$ equation alone and obtain an analytic effective EOM.

However, in the special case $d=0$, where the COM coincides with the multipole center, the rotational EOM decouples from the translational degrees of freedom and becomes a closed differential equation for $\theta$.
Hence, the mode-separation method can be applied to derive an effective EOM.
Below we focus on this $d=0$ case.

For $d=0$ and $T=0$, the rotational ULE is written as
\begin{align}
  &\ddot{\theta} = -\Gamma\dot{\theta} - B_{n,d=0}^{\mathrm{U}}\sin\qty(\omega t - \theta + \frac{\pi\qty(n-1)}{2n}),
  \label{eq:ULE-multipole-theta-eom-d=0}
\end{align}
with $\Gamma=\frac{2\gamma}{M}$ and $B_{n,d=0}^{\mathrm{U}}=\frac{4qE_{0}}{nM \ell}\csc\qty(\frac{\pi}{2n})$.
Substituting Eq.~\eqref{eq:multipole-mode-separation} into Eq.~\eqref{eq:ULE-multipole-theta-eom-d=0}, we obtain
\begin{align}
  \ddot{\theta}_{\mathrm{slow}} + \ddot{\theta}_{\mathrm{fast}} +
  \Gamma \qty(\dot{\theta}_{\mathrm{slow}} + \dot{\theta}_{\mathrm{fast}}) + B_{n,d=0}^{\mathrm{U}}\sin\qty(\omega t - \theta_{\mathrm{fast}} - \tilde{\theta}_{\mathrm{slow}}) = 0
  \label{eq:ULE-multipole-theta-separation}.
\end{align}
As the HF condition for mode separation, we assume
\begin{align}
  \epsilon_{\mathrm{U}} = \frac{B_{n,d=0}^{\mathrm{U}}}{\omega^{2}}\ll1\qquad\qty(\Leftrightarrow\ \omega\gg\sqrt{B_{n,d=0}^{\mathrm{U}}}). \label{eq:epsilon_U_def}
\end{align}
Similarly to the case of the OLE, we also assume that $\theta_{\mathrm{fast}}=\mathcal{O}(\epsilon_{\mathrm{U}})$ and $|\theta_{\mathrm{fast}}|\ll1$ hold.

Expanding Eq.~\eqref{eq:ULE-multipole-theta-separation} up to first order in $\theta_{\mathrm{fast}}$, we obtain
\begin{align}
  \ddot{\theta}_{\mathrm{slow}} + \ddot{\theta}_{\mathrm{fast}} + \Gamma \qty(\dot{\theta}_{\mathrm{slow}} + \dot{\theta}_{\mathrm{fast}}) + B_{n,d=0}^{\mathrm{U}}\qty[\sin\qty(\omega t - \tilde{\theta}_{\mathrm{slow}}) - \theta_{\mathrm{fast}}\cos\qty(\omega t - \tilde{\theta}_{\mathrm{slow}})] + \mathcal{O}\qty(\epsilon_{\mathrm{U}}^{3}\omega^{2}) = 0\label{eq:ULE-multipole-theta-separation-expanded}
\end{align}
Taking the time average of Eq.~\eqref{eq:ULE-multipole-theta-separation-expanded}, and using $\langle\ddot{\theta}_{\mathrm{fast}}\rangle_{\mathcal{T}} = \langle\dot{\theta}_{\mathrm{fast}}\rangle_{\mathcal{T}}=0$ and $\left\langle\sin\qty(\omega t - \tilde{\theta}_{\mathrm{slow}})\right\rangle_{\mathcal{T}} = 0$, we obtain
\begin{align}
  \ddot{\theta}_{\mathrm{slow}} + \Gamma \dot{\theta}_{\mathrm{slow}} - B_{n,d=0}^{\mathrm{U}}\left\langle\theta_{\mathrm{fast}}\cos\qty(\omega t - \tilde{\theta}_{\mathrm{slow}})\right\rangle_{\mathcal{T}}  + \mathcal{O}\qty(\epsilon_{\mathrm{U}}^{3}\omega^{2}) = 0\label{eq:ULE-multipole-theta-separation-averaged}.
\end{align}

In what follows, we expand $\theta_{\mathrm{fast}}=\theta_{\mathrm{fast}}^{(1)}+\mathcal{O}(\epsilon_{\mathrm{U}}^{2})$ with $\theta_{\mathrm{fast}}^{(k)}=\mathcal{O}(\epsilon_{\mathrm{U}}^{k})$.
From Eq.~\eqref{eq:ULE-multipole-theta-separation-averaged} we have $\ddot{\theta}_{\mathrm{slow}} + \Gamma \dot{\theta}_{\mathrm{slow}} + \mathcal{O}(\epsilon_{\mathrm{U}}^{2}\omega^{2}) = 0$, so the $\mathcal{O}(\epsilon_{\mathrm{U}}\omega^{2})$ terms in Eq.~\eqref{eq:ULE-multipole-theta-separation-expanded} yield
\begin{align}
  \ddot{\theta}_{\mathrm{fast}}^{(1)} + \Gamma\dot{\theta}_{\mathrm{fast}}^{(1)} + B_{n,d=0}^{\mathrm{U}}\sin\qty(\omega t - \tilde{\theta}_{\mathrm{slow}}) = 0 \label{eq:theta_f-1st-ULE}
\end{align}
Taking the steady periodic solution while neglecting the decaying homogeneous part, we obtain
\begin{align}
  &\theta_{\mathrm{fast}}^{(1)}=c_{1}\sin\qty(\omega t - \tilde{\theta}_{\mathrm{slow}}) + c_{2}\cos\qty(\omega t - \tilde{\theta}_{\mathrm{slow}}),
  \label{eq:theta-fast-(1)}
\end{align}
where $c_{1}=\frac{B_{n,d=0}^{\mathrm{U}}}{\omega^{2}+\Gamma^{2}}$ and $c_{2}=\frac{B_{n,d=0}^{\mathrm{U}}\Gamma}{\omega(\omega^{2}+\Gamma^{2})}$.
This confirms the condition $\theta_{\mathrm{fast}}^{(1)}=\mathcal{O}(\epsilon_{\mathrm{U}})$.
Substituting Eq.~\eqref{eq:theta-fast-(1)} into Eq.~\eqref{eq:ULE-multipole-theta-separation-averaged}, we obtain
\begin{align}
  &\ddot{\theta}_{\mathrm{slow}} + \Gamma \qty(\dot{\theta}_{\mathrm{slow}} - \Omega_{n}^{\mathrm{MS, U}}) + \mathcal{O}\qty(\epsilon_{\mathrm{U}}^{3}\omega^{2}) = 0 ,\label{eq:theta_s-1st-order-ULE}
\end{align}
where the parameter $\Omega_{n}^{\mathrm{MS, U}}$ is defined as
\begin{align}
  &\Omega_{n}^{\mathrm{MS, U}} = \frac{(B_{n,d=0}^{\mathrm{U}})^{2}}{2\omega(\omega^{2}+\Gamma^{2})} \label{eq:Omega_ULE_general}.
\end{align}
Therefore, in the NESS (long-time limit), the system relaxes to uniform rotation, $\dot{\theta}_{\mathrm{slow}} \simeq \Omega_{n}^{\mathrm{MS, U}} + \mathcal{O}(\epsilon_{\mathrm{U}}^{3}\omega)$.

In the weak-damping regime $\Gamma\ll\omega$, we have
\begin{align}
  \Omega_{n}^{\mathrm{MS, U}} &= \frac{(B_{n,d=0}^{\mathrm{U}})^{2}}{2\omega^{3}}\qty[1+\mathcal{O}\qty(\frac{\Gamma^{2}}{\omega^{2}})]
  = \Omega_{n}^{\mathrm{U}}\qty[1 + \mathcal{O}\qty(\frac{\Gamma^{2}}{\omega^{2}})]\label{eq:Omega_ULE_Gllw-ULE}
\end{align}
which is consistent with the effective ULE derived from the HF expansion in Floquet theory,
$\dot{\theta}\simeq\Omega_{n}^{\mathrm{U}}+\mathcal{O}(\omega^{-5})$.

On the other hand, in the strong-damping regime $\Gamma\gg\omega$, we obtain
\begin{align}
  \Omega_{n}^{\mathrm{MS, U}} = \frac{(B_{n,d=0}^{\mathrm{U}})^{2}}{2\omega\Gamma^{2}}\qty[1+\mathcal{O}\qty(\frac{\omega^{2}}{\Gamma^{2}})]
  = \Omega_{n}^{\mathrm{O}}\qty[1 + \mathcal{O}\qty(\frac{\omega^{2}}{\Gamma^{2}})]\label{eq:Omega_ULE_Gggw}.
\end{align}
In this limit, the inertial term in Eq.~\eqref{eq:theta_s-1st-order-ULE} is small, yielding the effective EOM
\begin{align}
  \dot{\theta}_{\mathrm{slow}} \simeq \Omega_{n}^{\mathrm{MS, U}} \simeq \Omega_{n}^{\mathrm{O}}
  \label{eq:effEOM_Gggw}.
\end{align}
Indeed, when $\Gamma$ is sufficiently large in the original ULE Eq.~\eqref{eq:ULE-multipole-theta-eom-d=0}, one may neglect the inertial term to obtain
\begin{align}
  \dot{\theta}\simeq -\frac{B_{n,d=0}^{\mathrm{U}}}{\Gamma}\sin\qty(\omega t-\theta+\frac{\pi(n-1)}{2n})
\end{align}
which has the same form as the OLE.
Namely, the mode separation method is consistent with the result of the HF expansion.

In summary, the mode-separation analysis for the restricted ULE case $d=0$ yields the general expressions Eqs.~\eqref{eq:theta_s-1st-order-ULE} and~\eqref{eq:Omega_ULE_general}.
Moreover, depending on the magnitude of $\frac{\Gamma}{\omega}=\frac{2\gamma}{M\omega}$, the effective ULE regime ($\Gamma\ll\omega$) and the effective OLE regime ($\Gamma\gg\omega$) are continuously connected through the NESS angular velocity $\dot{\theta}$.

Finally we again stress that it is not easy to derive the Floquet effective EOM of the ULE with $d\neq 0$ because $\bm R$ and $\theta$ are coupled with each other and the simple mode separation method for $\theta$ used in this section cannot be applied to the $d\neq 0$ case. On the other hand, the HF expansion of the Floquet theory is still powerful even for $d\neq 0$ and it implies that the HF expansion provides a systematic way of deriving the effective EOM for generic time-periodic classical systems.

%%%%%%%%%%%%%%%%%%%%%%%%%%%%%%%%%%%%%%
\subsection{Numerical Method: NESS and Floquet Rotation Frequency $\Omega$}\label{sec:NumericalMethod}
In the main text, we numerically computed the Floquet rotation frequency $\Omega$, defined as the time average of $\dot{\theta}$ after a sufficiently long time has elapsed since the laser irradiation was switched on for the electric multipolar model.
Here we explain how $\Omega$ is numerically determined in the nonequilibrium steady state (NESS).
%%%%%%%%%%%%%%%%%%%%%%%%%%%%%%%%%%%%%
\begin{figure}[tb]
  \centering
  \includegraphics[width=0.7\textwidth]{"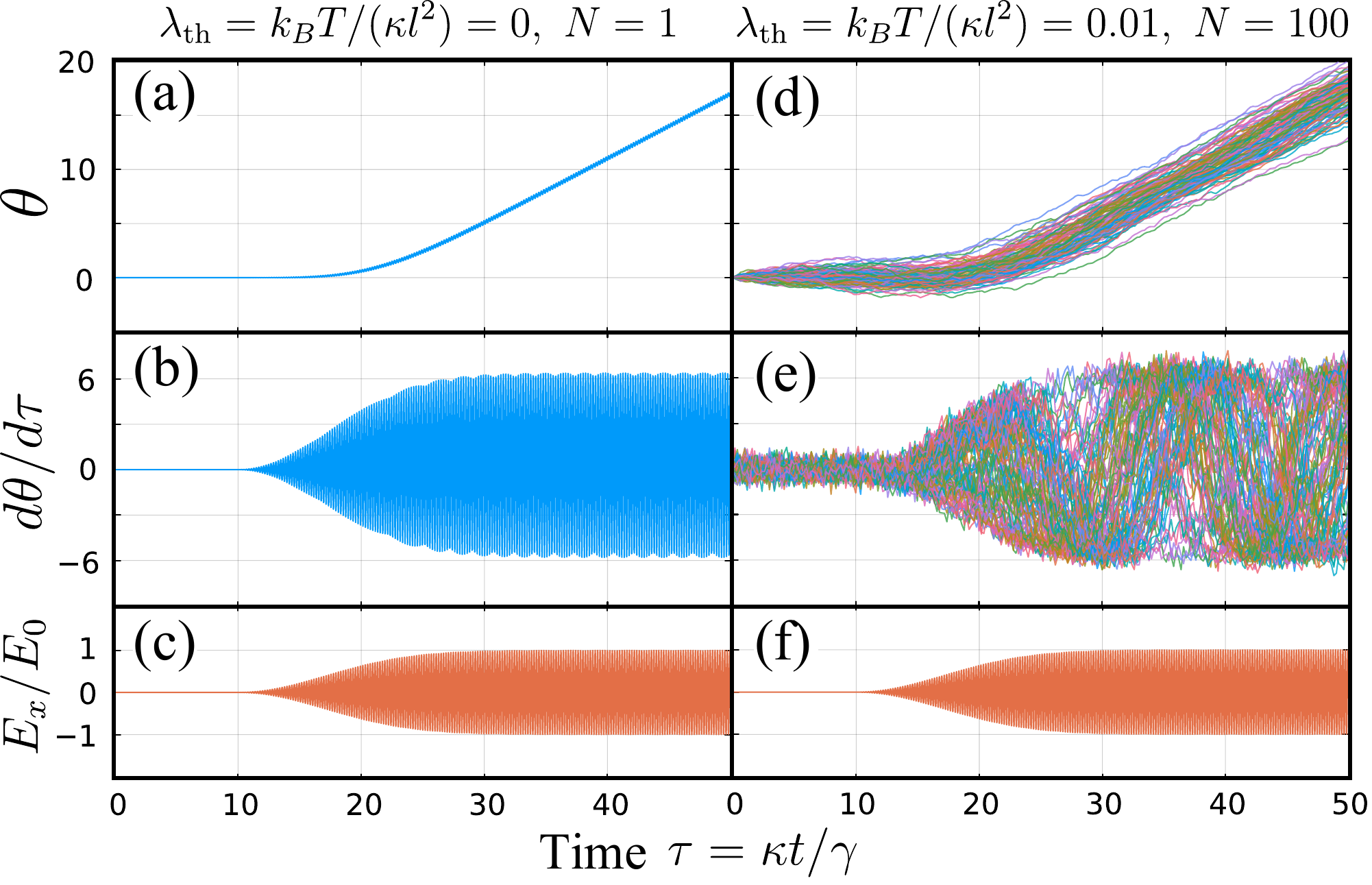"}
  \caption{
    Time evolution of rotational motion in the electric dipolar model with $m_1=m_2$.
    The left column (a--c) shows the results at absolute zero ($\lambda_{\mathrm{th}}=0$, $N=1$), and the right column (d--f) shows the finite-temperature results ($\lambda_{\mathrm{th}}=0.01$, $N=100$).
    (a,d) show the rotation angle $\theta$, (b,e) show the angular velocity $d\theta/d\tau$, and (c,f) show the time evolution of the $x$ component of the laser electric field, $E_x/E_0$.
    The horizontal axis is the dimensionless time $\tau=\kappa t/\gamma$, and the laser is turned on at $\tau=10$.
    The parameters are $\lambda_{\mathrm{m}}=10^{-0.7},\ \lambda_{\mathrm{fre}}=10^{1.5},\ \lambda_{\mathrm{el}}=10^{1.0}$; for the finite-temperature case, $N=100$ trajectories are overlaid.
  }
  \label{fig:NESS}
\end{figure}
%%%%%%%%%%%%%%%%%%%%%%%%%%%%%%%%%%%%%

When the laser field is switched on at $\tau=10$ [Fig.~\ref{fig:NESS}(c)], the system passes through a transient regime and then a fast oscillatory component appears in the angular velocity $\dot{\theta}$, originating from the linear response to the external field [Fig.~\ref{fig:NESS}(b)].
On the other hand, as is clear from Fig.~\ref{fig:NESS}(a), on a slow time scale the time evolution of $\theta$ increases almost linearly, so that the motion can be regarded as uniform rotation on average.
In this study, we define this regime as a NESS, and determine the rotational frequency $\Omega$ from the slope of $\theta$, obtained by a linear fit over a finite time window $\Delta\tau$ within the NESS.
Here $\Delta\tau$ is chosen to be much longer than the driving period to average out fast oscillations.

For finite-temperature calculations, we introduce the dimensionless parameter $\lambda_{\mathrm{th}} = k_{B}T/(\kappa \ell^{2})$ [see Eq.~\eqref{eq:lam_th}].
Figures~\ref{fig:NESS} (d--f) show a typical time evolution for the dipolar model at a finite temperature of $\lambda_{\mathrm{th}}=0.01$.
At finite temperatures ($\lambda_{\rm th}>0$), the random force $\bm h(t)$ introduces thermal fluctuations to $\theta$ and $\dot{\theta}$, as seen in Panels (d) and (e).
In this case, we first compute $\Omega$ for each realization from a time-moving-window average over an interval after reaching the NESS, and then take an ensemble average to reduce the statistical error; we denote this averaged value by $\langle\Omega\rangle$.
Here, $\langle\cdots\rangle$ denotes the ensemble average over $N$ independent noise
realizations, i.e., $\langle \Omega \rangle = \frac{1}{N}\sum_{s=1}^{N}\Omega_s$,
where $\Omega_s$ is the rotation frequency obtained from the $s$-th time evolution. For more detail, see Sec.~\ref{sec:thermal} for the finite-temperature numerical results.

%%%%%%%%%%%%%%%%%%%%%%%%%%%%%%%%%%%%%%%%%%%%%
\subsection{New Phase in Mass Imbalance Case of the Underdamped Dipolar Model}
In the main text, we verified the mass ratio dependence of the Floquet rotation in the ULE with a mass imbalance, while we did not discuss the ``phase'' boundary between two regimes with different exponents $\alpha$.
In this section, we briefly discuss such a phase boundary in the mass-imbalance case and find a novel phase in the NESS.

\begin{figure}[tb]
  \centering
  \includegraphics[width=0.8\linewidth]{"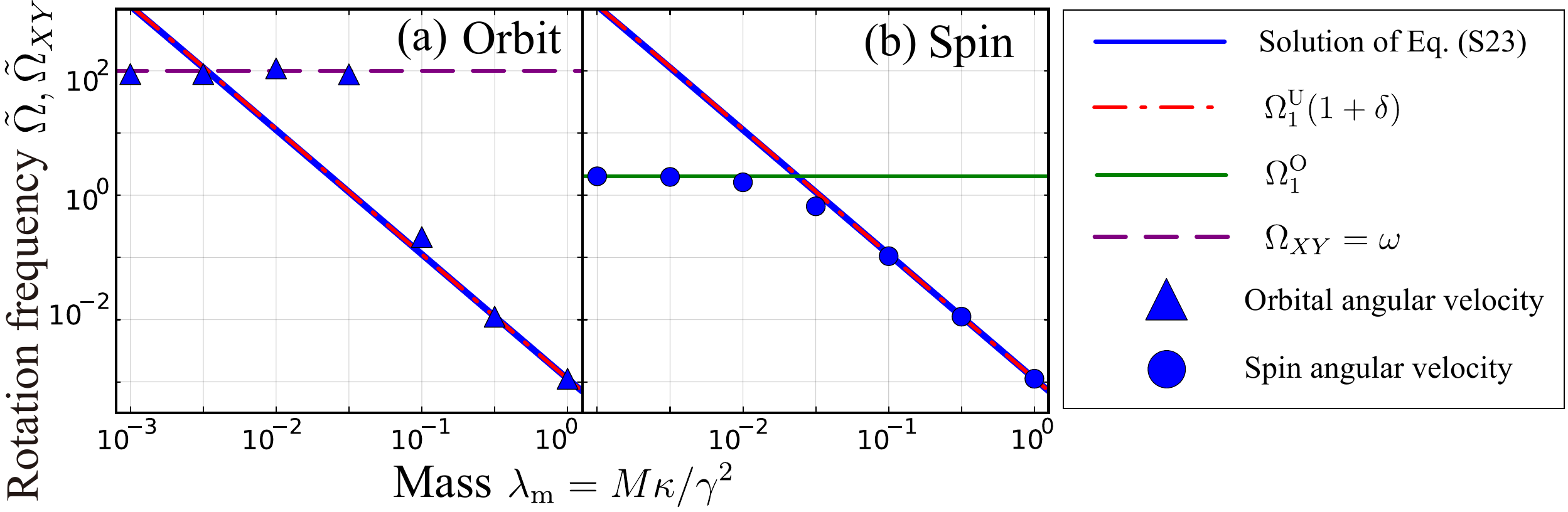"}
  \caption{
    Panels (a) and (b) show, for $m_{1}/m_{2}=0.5$, the dependence on the total-mass parameter $\lambda_{\mathrm{m}}$ (horizontal axis) of the (a) orbital rotation frequency $\tilde{\Omega}_{XY}=\gamma\Omega_{XY}/\kappa$ and (b) spin rotation frequency $\tilde{\Omega}=\gamma\Omega/\kappa$ (vertical axis), respectively.
    Points (triangles and circles) are the numerical results of the ULE.
    The blue solid, red dash-dotted, and green solid lines represent the numerical solution of Eq.~\eqref{eq:dipole-Omega-equation} for $\Omega_{\mathrm{rot}}$, its approximation $\Omega_{1}^{\mathrm{U}}(1+\delta)$ from Eq.~\eqref{eq:Delta}, and $\Omega_{1}^{\mathrm{O}}$, respectively.
    The purple dashed line in panel (a) indicates the drive frequency $\Omega_{XY}=\omega$, which the orbital rotation approaches in the small-$\lambda_{\mathrm{m}}$ regime, where the slow Floquet orbital rotation disappears.
    These results are obtained for the dipolar model with $\lambda_{\mathrm{fre}}=10^{2}$, $\lambda_{\mathrm{el}}=10$, and $\lambda_{\mathrm{th}}=0$.
  }
  \label{fig:mass-ratio}
\end{figure}

Panels (a) and (b) of Fig.~\ref{fig:mass-ratio} show the dependence on the total-mass parameter $\lambda_{\mathrm{m}} = M\kappa/\gamma^2$ with a fixed mass ratio $m_{1}/m_{2}=0.5$.
For the spin rotation in panel (b), in the large-$\lambda_{\mathrm{m}}$
regime, the numerical ULE result is well reproduced by the steady-state
solution $\Omega_{\mathrm{rot}}$ ($\alpha=-3$) of Eq.~\eqref{eq:dipole-Omega-equation}, which is derived
from the effective ULE for the mass-imbalance case.
As also shown in Fig.~5 of the main text for the mass-ratio dependence,
the approximate expression $\Omega_{1}^{\mathrm{U}}(1+\delta)$ provides
a good approximation to this solution.
As $\lambda_{\mathrm{m}}$ decreases, the numerical result approaches
the effective OLE result $\Omega_{1}^{\mathrm{O}}$ ($\alpha=-1$).
In panel (a), the orbital rotation in the large-$\lambda_{\mathrm{m}}$
regime is likewise well described by the solution of Eq.~\eqref{eq:dipole-Omega-equation} with $\alpha=-3$.
In contrast, in the small-$\lambda_{\mathrm{m}}$ regime, the slow
Floquet orbital rotation disappears and the numerical result approaches
$\Omega_{XY}=\omega$ with $\alpha=1$.

From these results we can conclude that in the mass-imbalance case, there is a new ``phase'' where the spin rotation obeys $\alpha=-1$ and the orbital rotation does $\alpha=1$ [see the small-$\lambda_{\mathrm{m}}$ regime in Panels (a) and (b)].
In other words, the COM coordinate follows the ac electric field, while the autorotation follows the Floquet rotation.

%%%%%%%%%%%%%%%%%%%%%%%%%%%%%%%%%%%%%
%%%%%%%%%%%%%%%%%%%%%%%%%%%%%%%%%%%%%

%%%%%%%%%%%%%%%%%%%%%%%%%%%%%%%%%%%%%%%%%%%%%%%
\subsection{Damping Dependence of Floquet Rotation}
%%%%%%%%%%%%%%%%%%%%%%%%%%%%%%%%%%%%%
\begin{figure}[]
  \centering
  \includegraphics[width=0.5\linewidth]{"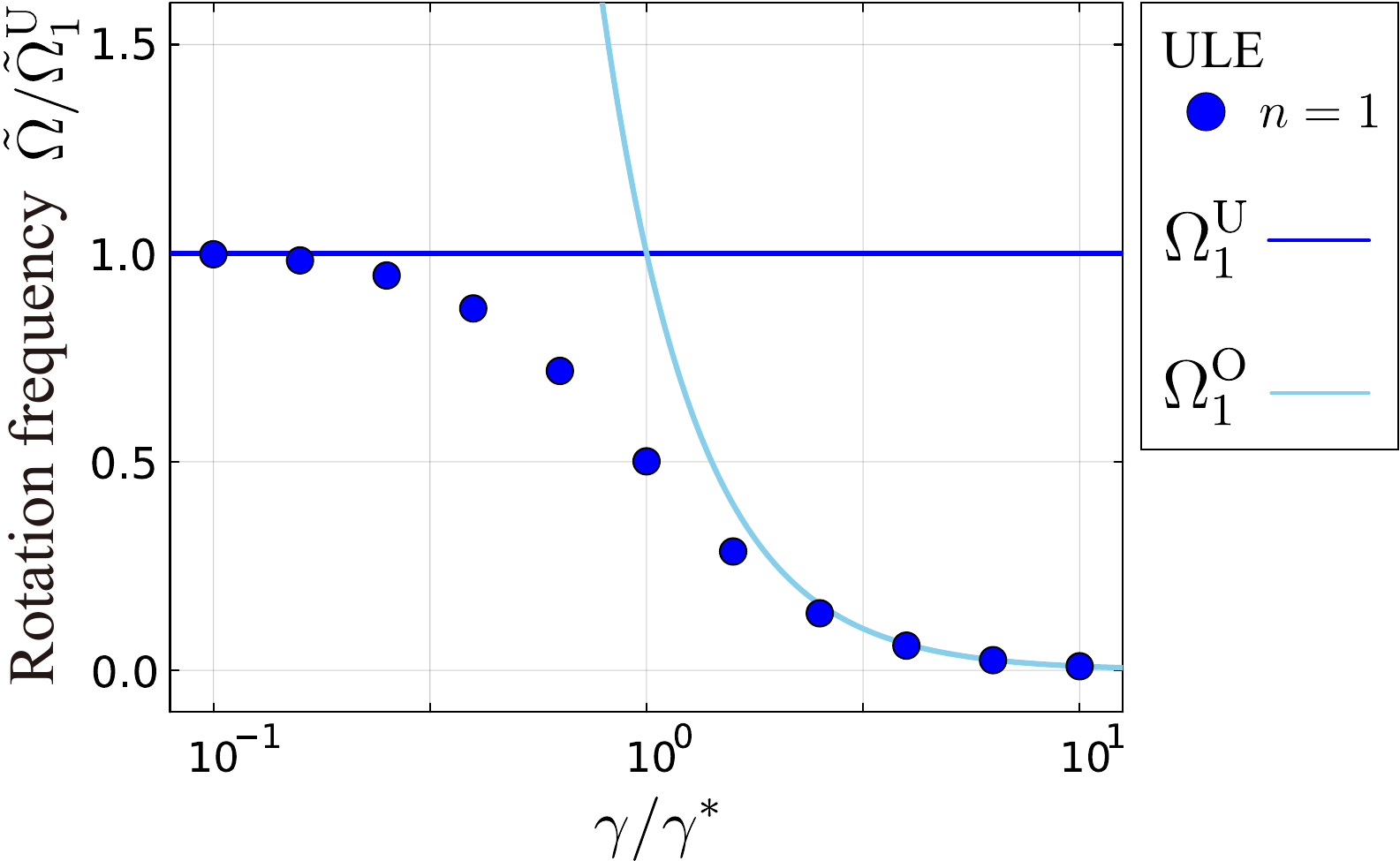"}
  \caption{
    Dependence of the rotational frequency on the drag coefficient $\gamma$ for the equal-mass electric dipolar model ($n=1$, $m_{1}=m_{2}$) at $T=0$.
    The horizontal axis is the normalized drag coefficient $\gamma/\gamma^{*}$, where $\gamma^{*}\coloneq \frac{M\omega}{2}$.
    The vertical axis shows the rotational frequency normalized by $\Omega_{1}^{\mathrm{U}}$ obtained from the effective ULE, i.e., $\Omega/\Omega_{1}^{\mathrm{U}}=\tilde{\Omega}/\tilde{\Omega}_{1}^{\mathrm{U}}$.
    The markers represent the numerical results of the ULE.
    The blue and light-blue solid lines denote $\Omega_{1}^{\mathrm{U}}$ and $\Omega_{1}^{\mathrm{O}}\propto\gamma^{-2}$, respectively.
    The results are obtained for $\lambda_{\mathrm{m}}^{*}=0.04$, $\lambda_{\mathrm{fre}}^{*}=50$, $\lambda_{\mathrm{el}}=2$, and $\lambda_{\mathrm{th}}=0$.
  }
  \label{fig:gamma}
\end{figure}
%%%%%%%%%%%%%%%%%%%%%%%%%%%%%%%%%%%%%
In this section, we discuss the dependence of the Floquet rotation frequency on the drag coefficient $\gamma$.
The paper of Ref.~\cite{reimann2018ghz} has reported that the Floquet rotation frequency of a silica nanoparticle increases as the environment-gas pressure is reduced.
If the drag coefficient $\gamma$ is regarded as proportional to the gas pressure, this observation suggests a $\gamma$ dependence $\Omega\propto\gamma^{-1}$.
On the other hand, the analytical results obtained from the HF expansion yield the limiting drag dependences $\Omega^{\mathrm{U}}_{n}\propto\gamma^{0}$ for the effective ULE and $\Omega^{\mathrm{O}}_{n}\propto\gamma^{-2}$ for the effective OLE.

To isolate the dependence on the drag coefficient, we consider varying only $\gamma$ while keeping $M$, $\omega$, $\kappa$, $E_{0}$, and $\ell$ fixed.
We define the characteristic drag coefficient as $\gamma^{*} = \frac{M\omega}{2}$.
Here, we denote $\lambda_{\mathrm{m}}^{*}=M\kappa/\gamma^{*2}$ and $\lambda_{\mathrm{fre}}^{*}=\gamma^{*}\omega/\kappa$.
Figure~\ref{fig:gamma} shows the numerical results for the dipolar model described by the ULE at $T=0$.
The normalized rotation frequency $\tilde{\Omega}/\tilde{\Omega}_{1}^{\mathrm{U}}$ smoothly interpolates between the effective ULE plateau $\tilde{\Omega}/\tilde{\Omega}_{1}^{\mathrm{U}}=1$ and the effective OLE behavior $\tilde{\Omega}=\tilde{\Omega}_{1}^{\mathrm{O}}\propto\gamma^{-2}$.

This figure implies that our dipolar model does not reproduce the experimentally suggested scaling $\Omega\propto\gamma^{-1}$ as a broad asymptotic regime.
On the other hand, in the crossover region between the effective ULE and the effective OLE, an intermediate $\gamma$ dependence appears, including a region where the rotational frequency approximately agrees with $\Omega\propto\gamma^{-1}$.
Thus, the result of Fig.~\ref{fig:gamma} suggests that at least part of the experimentally observed $\gamma^{-1}$ trend may originate from the ULE--OLE crossover, while at the same time indicating that additional mechanisms omitted in our minimal model may be important in explaining the broad experimental window.

%%%%%%%%%%%%%%%%%%%%%%%%%%%%%%%%%%%%%
\subsection{Temperature Effects on Floquet Rotation}\label{sec:thermal}
%%%%%%%%%%%%%%%%%%%%%%%%%%%%%%%%%%%%%
\begin{figure}[]
  \centering
  \includegraphics[width=0.8\linewidth]{"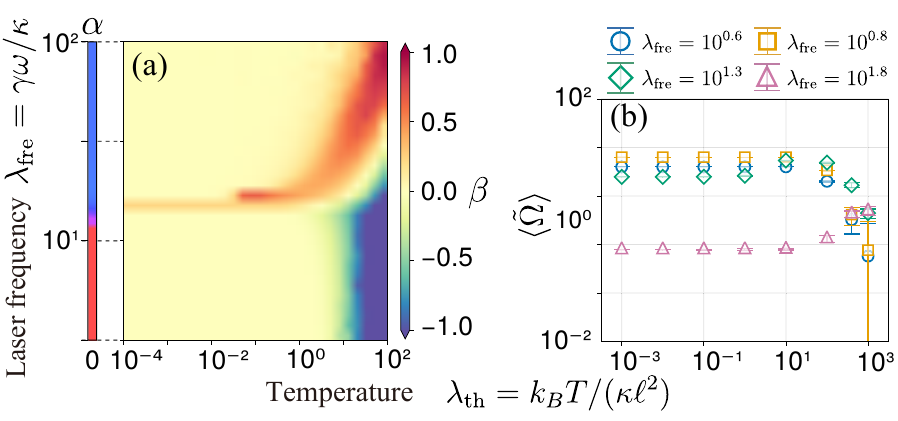"}
  \caption{
    Temperature dependence of the rotation frequency in the ULE of the dipolar model.
    The horizontal axis is the temperature $\lambda_{\mathrm{th}}$ in both panels (a) and (b).
    Panel (a) shows the color plot of the value of $\beta$ in the space $(\lambda_{\mathrm{th}}, \lambda_{\mathrm{fre}})$.
    The parameter $\beta$ is defined in Eq.~\eqref{eq:def_beta} and stands for a log-scaled normalized rotation frequency.
    The band at $\lambda_{\mathrm{th}}=0$ on the left shows the value of $\alpha$ as a function of the laser frequency $\lambda_{\mathrm{fre}}$: Red and blue regimes respectively correspond to $\alpha=1$ and $-3$. The range of vertical axis $\lambda_{\mathrm{fre}}=10^{0.5}\text{-}10^{2}$ is a part of Fig.~4(e) at $\lambda_{\mathrm{m}}=0.2$.
    In panel (b), we extract cuts from (a) at $\lambda_{\mathrm{fre}}=10^{0.6},10^{0.8},10^{1.3},10^{1.8}$ and plot them with the vertical axis taken as the rotation frequency $\langle\tilde{\Omega}\rangle$.
    The results are obtained for $\lambda_{\mathrm{el}}=10$ with ensemble size $N=\mathcal{O}(10^{2}\text{-}10^{6})$, with sampling continued until sufficient convergence is achieved; error bars indicate $\pm 3\sigma$ ($\sigma$ is the deviation).
  }
  \label{fig:thermal-ule}
\end{figure}
%%%%%%%%%%%%%%%%%%%%%%%%%%%%%%%%%%%%%
%%%%%%%%%%%%%%%%%%%%%%%%%%%%%%%%%%%%%
\begin{figure}[]
  \centering
  \includegraphics[width=0.8\linewidth]{"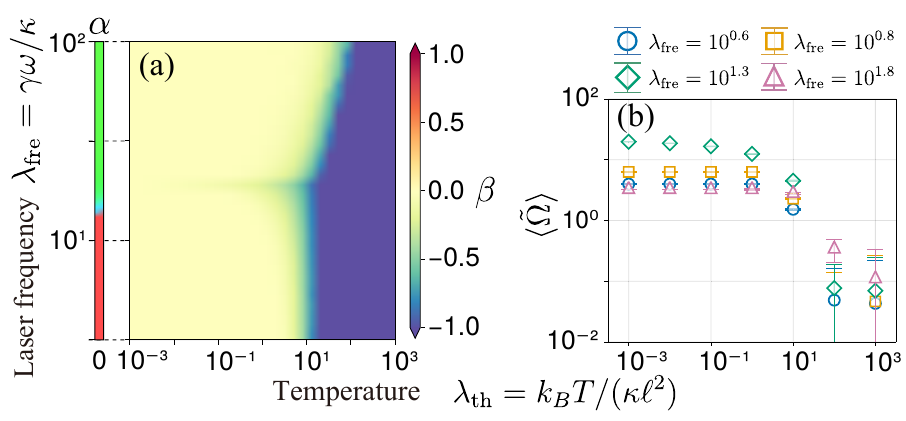"}
  \caption{
    Temperature dependence of the rotation frequency in the OLE of the dipolar model.
    The horizontal axis is the temperature $\lambda_{\mathrm{th}}$ in both panels (a) and (b).
    Panel (a) shows the color plot of the value of $\beta$ in the space $(\lambda_{\mathrm{th}}, \lambda_{\mathrm{fre}})$.
    The parameter $\beta$ is defined in Eq.~\eqref{eq:def_beta} and stands for a log-scaled normalized rotation frequency.
    The band at $\lambda_{\mathrm{th}}=0$ on the left shows the value of $\alpha$ as a function of the laser frequency $\lambda_{\mathrm{fre}}$: Red and green regimes respectively correspond to $\alpha=1$ and $-1$. The range of vertical axis $\lambda_{\mathrm{fre}}=10^{0.5}\text{-}10^{2}$ is a part of Fig.~3(c) at $\lambda_{\mathrm{el}}=10$.
    In panel (b), we extract cuts from (a) at $\lambda_{\mathrm{fre}}=10^{0.6},10^{0.8},10^{1.3},10^{1.8}$ and plot them with the vertical axis taken as the rotation frequency $\langle\tilde{\Omega}\rangle$.
    The ensemble averages are taken over $N=\mathcal{O}(10^{2}\text{--}10^{4})$ realizations, with sampling continued until sufficient convergence is achieved; error bars indicate $\pm 3\sigma$ ($\sigma$ is the standard deviation).
  }
  \label{fig:thermal-ole}
\end{figure}
%%%%%%%%%%%%%%%%%%%%%%%%%%%%%%%%%%%%%
So far we have focused mainly on the zero-temperature limit. In this section, we quantitatively discuss how a finite temperature affects the Floquet rotation in the NESS.

Figures~\ref{fig:thermal-ule} and \ref{fig:thermal-ole} show the temperature dependence of the rotation frequency in the dipolar model for the ULE and OLE, respectively.
The parameter $\beta$ represented by the color in panels (a) of both figures is defined as
\begin{align}
  \beta \coloneq \log_{10}\qty(\langle\Omega\rangle/\Omega_{T=0}), \label{eq:def_beta}
\end{align}
where $\Omega_{T=0}$ is the rotation frequency at absolute zero.
Thus, $\beta=0$ indicates that the rotation frequency is unchanged from its $T=0$ value, while $\beta<0$ ($\beta>0$) indicates suppression (enhancement) of the rotation by thermal fluctuations.
In both cases of the ULE and OLE, the rotation frequency remains nearly unchanged from its zero-temperature value in a broad low-temperature regime $\lambda_{\mathrm{th}}\lesssim10^{1}$ [the light yellow regimes in panels (a) of Figs.~\ref{fig:thermal-ule} and \ref{fig:thermal-ole}].
Thus, thermal fluctuations have little effect on the rotational motion in this low-temperature regime.
As $\lambda_{\mathrm{th}}$ is further increased, deviations from the zero-temperature response become appreciable, indicating that thermal fluctuations begin to affect the rotational dynamics.

Let us estimate the temperature $T$ by using $\lambda_{\mathrm{th}}\sim 10^{-2}$ and the typical values of $\kappa$ and $\ell$.
When we set $\kappa=\Unit{10^{-5}}{kg/s^{2}}$ and $\ell=\Unit{100}{nm}$,
$\lambda_{\mathrm{th}}\simeq4\times10^{-2}$ ($\simeq6\times10^{-2}$) corresponds to $T=\Unit{300}{K}$ ($T=\Unit{400}{K}$).
This means that room temperature is much lower than the boundary of the low- and high-temperature regimes, $\lambda_{\mathrm{th}}\sim 10^{1}$.
Moreover, the experimentally relevant laser-frequency parameter is expected to be much larger than the range considered here.
Using the typical values $\gamma=\Unit{10^{-9}}{kg/s}$, $\kappa=\Unit{10^{-5}}{kg/s^{2}}$, and $\omega=\Unit{2\times10^{15}}{rad/s}$, we estimate $\lambda_{\mathrm{fre}}\simeq2\times10^{11}$.
Since the boundary between the low- and high-temperature regimes shifts toward larger $\lambda_{\mathrm{th}}$ with increasing $\lambda_{\mathrm{fre}}$ in the Floquet-rotation regime, this trend suggests that thermal fluctuations should be even less relevant under experimentally realistic conditions.
Therefore, for experimentally relevant temperatures around room temperature, the effect of thermal fluctuations on the Floquet rotation is expected to be negligible within the present model.

%%%%%%%%%%%%%%%%%%%%%%%%%%%%%%%%%%%%%%%%
\subsection{Movies}
In this section, we describe the contents of the Supplemental Movies.
All movies show time-evolution sequences obtained by numerically solving the EOM of the dipolar model.
In each panel, the arrow at the upper right indicates the applied optical electric-field vector.

In Movie~1, the top row (a) [(c)] shows the solution of the EOM for the dipolar model under right-circularly polarized light (RCPL) with $\lambda_{\mathrm{fre}}=10$ [left-circularly polarized light (LCPL) with $\lambda_{\mathrm{fre}}=-10$], with parameters $\lambda_{\mathrm{m}}=0.02,\ \lambda_{\mathrm{el}}=2,\ \lambda_{\mathrm{th}}=0$.
In this case, the system rotates on average in one direction on a slow time scale while exhibiting a fast oscillatory component originating from the linear response to the electric field.
The bottom row (b) [(d)] shows the time evolution obtained by solving the effective EOM derived from the Floquet theory for (a) [(c)].
This confirms that only the slow uniform-rotation component corresponding to Floquet rotation is extracted.
As a technical remark, panels (a) and (c) are obtained by solving the ULE.
However, for these parameters, the inertial effect is sufficiently weak that the dynamics is well described by the effective OLE.
Therefore, panels (b) and (d) are obtained by solving the effective OLE.

Next, for Movie~2, panels (a,b) are the same as panels (a,b) in Movie~1, while panels (c,d) show the results of simulations for the quadrupolar model using the same parameters as in (a,b).
Panel (c) shows the solution of the EOM under RCPL with $\lambda_{\mathrm{fre}}=-10$, with $\lambda_{\mathrm{m}}=0.02,\ \lambda_{\mathrm{el}}=2,\ \lambda_{\mathrm{th}}=0$.
Panel (d) shows the time evolution obtained by solving the effective EOM derived from Floquet theory for panel (c).
Because the Floquet-rotation coefficient is renormalized, the angular velocity of the slow uniform-rotation component becomes smaller than that of the dipolar model in (a,b).
As above, (c) is obtained by solving the ULE, whereas (d) is obtained by solving the effective OLE.

In Movie~3, panel (a) [(b)] shows the solution of the EOM for the electric dipolar model under RCPL with $\lambda_{\mathrm{fre}}=2$ [LCPL with $\lambda_{\mathrm{fre}}=-2$], with parameters
$\lambda_{\mathrm{m}}=0.02,\ \lambda_{\mathrm{el}}=4,\ \lambda_{\mathrm{th}}=0$.
In this case, the conditions required for the HF expansion of the OLE are not satisfied, and the HF description breaks down.
As a result, one observes a rotational motion in which the dipolar model follows the optical electric field with $\Omega=\omega$.

\end{document}